\documentclass[%
 reprint,
 amsmath,amssymb,
 aps,
floatfix,
]{revtex4-2}

\usepackage{graphicx}
\usepackage{dcolumn}
\usepackage{bm}
\usepackage{physics}
\usepackage{multirow}

\usepackage{color,soul}
\usepackage{xspace}
\usepackage{xcolor}
\usepackage{siunitx}

\begin{document}

\preprint{APS/123-QED}

\title{Modeling Fluid Polyamorphism Through a Maximum-Valence Approach}

\author{Nikolay A. Shumovskyi}%
\affiliation{Department of Physics, Boston University, Boston, MA 02215, USA}

\author{Thomas J. Longo}%
\affiliation{Institute for Physical Science and Technology, University of Maryland, College Park, MD 20742, USA}

\author{Sergey V. Buldyrev}%
\email{buldyrev@yu.edu}
\affiliation{ Department of Physics, Yeshiva University, New York, NY 10033, USA    \\                  Department of Physics, Boston University, MA 02215, USA}

\author{Mikhail A. Anisimov}%
\affiliation{Department of Chemical and Biomolecular Engineering and Institute for Physical Science
and Technology, University of Maryland, College Park, MD 20742, USA}

\date{\today}

\begin{abstract}
We suggest a simple model to describe polyamorphism in single-component fluids using a maximum-valence approach. The model contains three types of interactions:  i) atoms attract each other by van der Waals forces that generate a liquid-gas transition at low pressures, ii) atoms may form covalent bonds that induce association, and iii) bonded atoms attract or repel  each other stronger than non-bonded atoms, thus generating liquid-liquid separation. As an example, we qualitatively compare this model with the behavior of liquid sulfur and show that condition (iii) {generates} a liquid-liquid phase transition in addition to the liquid-gas phase transition.
\end{abstract}

\maketitle


The existence of two alternative liquid phases in a single-component substance is known as ``liquid polyamorphism'' \cite{Stanley_Liquid_2013,Anisimov2018,Tanaka_Liquid_2020}. A substance may be found to be polyamorphic by experimentally or computationally detecting a liquid-liquid phase transition (LLPT), which could be terminated at a liquid-liquid critical point (LLCP) \cite{Franzese2001,Sciorino_Silicon_2011}. Liquid polyamorphism has been observed in a variety of substances including: hydrogen { \cite{Ohta_H_2015,McWilliams_H_2016,Zaghoo_H_2013,Zaghoo_H_2016,Zaghoo_H_2017}}, helium \cite{Vollhardt_He_1990,Schmitt_He_2015}, sulfur \cite{Henry2020}, phosphorous \cite{Katayama2000,Katayama_Phos2_2004} and liquid carbon \cite{Glosli_Liquid_1999}, while being proposed to exist in selenium and tellurium \cite{Brazhkin_PT_1999,Plasienka_Structural_2015} {and in various oxides \cite{Tanaka_Liquid_2020} - \textit{e.g.} silica \cite{Saika_Silica_2004,Lascaris_Silica_2014,Chen_Silica_2017}}. It has also been hypothesized in supercooled {silicon \cite{Sastry_Silicon_2003,Sciorino_Silicon_2011} and in metastable deeply supercooled} water below the temperature of spontaneous ice nucleation \cite{Stanley_Liquid_2013,Anisimov2018,Tanaka_Liquid_2020,Holten_Liquid_2012,Gallo2016,Duska_Water_2020,Caupin_Thermodynamics_2019,Poole1992,Holten2001,Debenedetti2020,Biddle_Two_2017,Debenedetti_One_1998}.

The phenomenon of liquid polyamorphism could be understood through the interconversion of the two alternative molecular or supramolecular states via a reversible reaction \cite{Anisimov2018,Longo2021,Caupin2021}. While for some polyamorphic systems, like supercooled water, this approach is still being debated, there are substances (such as hydrogen, sulfur, phosphorous, and liquid carbon) where liquid-liquid phase separation is indeed induced by a chemical reaction. For example, it was recently discovered that high-density sulfur, well above the liquid-gas critical pressure (in the range from $0.5$ to $\SI{2.0}{\giga\pascal}$), exhibits a LLPT indicated by a discontinuity in density from a low-density-liquid (LDL) monomer-rich phase to a high-density-liquid (HDL) polymer-rich phase \cite{Henry2020}. This liquid-liquid transition is found in a polymerized state of sulfur (observed above $\SI{160}{\degreeCelsius}$ at ambient pressure \cite{Sauer_Lambda_1967,Bellissent_Sulfur_1994,Kozhevnikov_Sulfur_2004,Tobolsky_Sulfur_1959,Tobolsky_Selenium_1960}). However, with further increase of temperature, as the system approaches the liquid-gas phase transition (LGPT), the polymer chains gradually dissociate. Another liquid-liquid transition accompanied by a {chemical} reaction has been {predicted} in hydrogen at extremely high-pressures \cite{Morales_H_2010,Pierleoni_H_2016,Geng_H_2019,Heinz_H_2020,Cheng_H_2020}, and {although the first-order phase transition is still a subject of debate in the literature \cite{Comment_Zaghoo_2016_1,Comment_Zaghoo_2016_2,Reply_Comment_Zaghoo_2016}, two liquid phases of hydrogen have been observed}, in which liquid-molecular hydrogen (dimers) dissociates into atomistic-metallic hydrogen \cite{Ohta_H_2015,McWilliams_H_2016,Zaghoo_H_2013,Zaghoo_H_2016,Zaghoo_H_2017}.

In this work, motivated by the recent discoveries of the LLPT in sulfur \cite{Henry2020} and hydrogen {\cite{Ohta_H_2015,McWilliams_H_2016,Zaghoo_H_2013,Zaghoo_H_2016,Zaghoo_H_2017}}, we propose a simple model to describe liquid polyamorphism in a variety of chemically-reacting fluids. The model combines the ideas of two-state thermodynamics \cite{Anisimov2018,Holten2001} with the maximum-valence approach \cite{Zaccarelli2005,Debenedetti,Debenedetti2}, in which atoms may form covalent bonds via a reversible reaction, changing their state according to their bond number. By mimicking the valence structure and maximum bond number, $z$, our model predicts the LLPT in systems with dimerization ($z=1$), polymerization ($z=2$), and gelation ($z>2$). As an example, we compare the molecular dynamics (MD) simulations with the phase behavior of sulfur. In particular, we show that when the bonded atoms attract each other stronger than to the unbonded atoms, phase separation is coupled to polymerization generating the LLPT in sulfur.

\begin{figure*}[t]
\centering
{
\includegraphics[width=.7\textwidth]{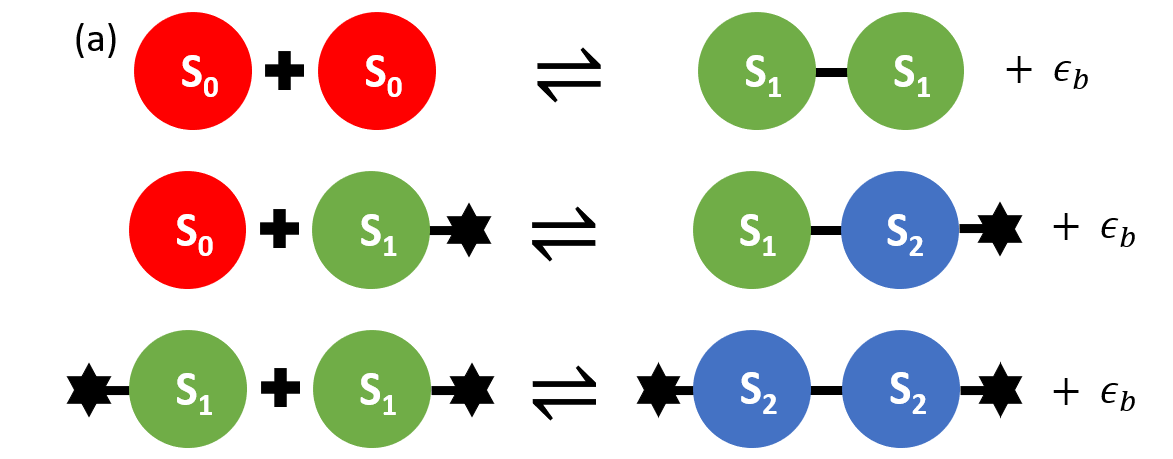}
}
\includegraphics[width=0.32\linewidth]{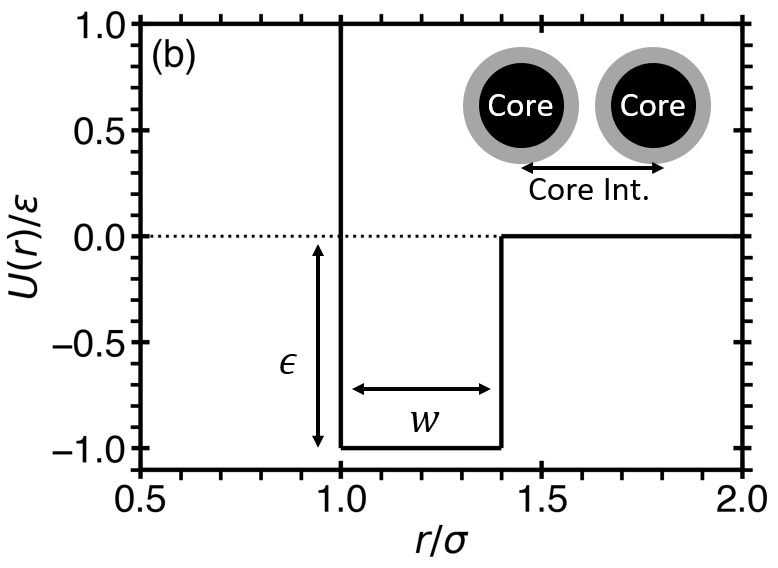}
\includegraphics[width=0.32\linewidth]{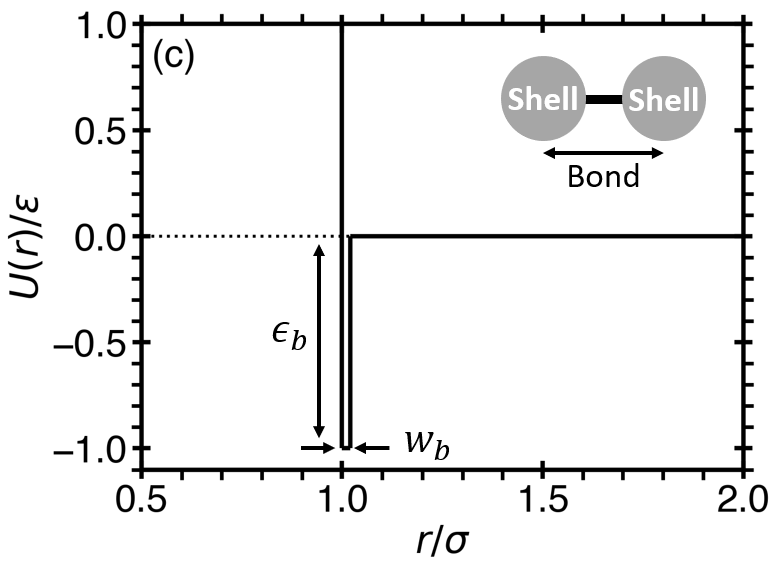}
\includegraphics[width=0.32\linewidth]{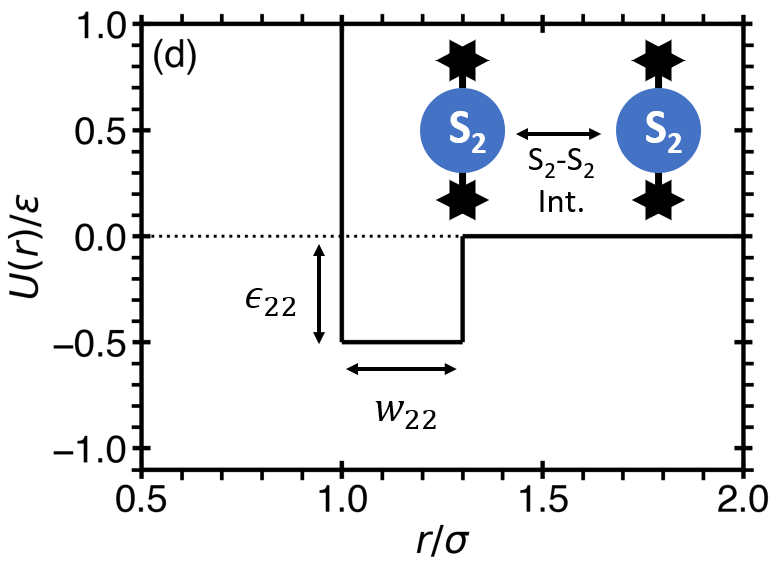}
\caption{Reactions and interactions in the maximum-valence model. (a) The three types of covalent bond-forming reversible chemical reactions that may occur in the system. If two atoms without bonds (S$_0$) collide with each other, they may form a bond and become S$_1$ atoms. If a S$_0$ and S$_1$ atom collide, they may form a bond and become S$_1$ and S$_2$ atoms, respectively. If two S$_1$ atoms collide with each other, they form an additional bond and become S$_2$ atoms. (b-d) The three major interactions between atoms, in which each atom is composed of a core and shell, both with a radius $\sigma$ and mass $m$. $U(r)$ is the pair potential energy and $r$ is the distance from the center of an atom. (b) The cores of each atom interact with an attractive square well of depth $\epsilon=1$ and width $w=0.4$. (c) The shells may react to form covalent bonds that consist of a narrow well with depth $\epsilon_{b} = 1$ and width $w_{b}=0.02$. (d) Phase segregation is coupled to polymerization via the additional attractive interactions between atoms in state S$_2$, described by a square well of depth $\epsilon_{22} = 0.5$ and width $w_{22}=0.3$.
}
\label{fig1}
\end{figure*}

\section{Maximum-Valence Model} We model the polymerization of a sulfur-like system $(z=2)$ by characterizing each atom by its coordination number, the number of bonds it has with other atoms. Depending on the coordination number, each atom is assigned to distinguished states: S$_0$ (with zero bonds), S$_1$ (with one bond), and S$_2$ (with two bonds). Atoms cannot form more than two bonds and, consequently, will polymerize into a linear polymer. All of the atoms in the system may change their state by forming or breaking a covalent bond via a reversible reaction. Fig.~\ref{fig1}a depicts the three types of reversible reactions that may occur in the system. In this work, we demonstrate that the minimum ingredients required to produce a LLPT are the following: i) the van der Waals interactions between atoms, which produce a LGPT; ii) covalent bonds between atoms, which induce polymerization; and iii), as we hypothesize, additional van der Waals interactions between atoms with maximum valency (having two bonds), which couple phase segregation to polymerization. These three ingredients are illustrated by square-well potentials in Figs.~\ref{fig1}(b-d). 

Physically, the additional attraction between atoms in neighboring chains may stem from the fact that in real polymers the covalent bond is shorter than the diameter of the unbonded (``free'') atoms, such that the attractive wells of bonded atoms in neighboring chains overlap with each other \cite{Stell1972,Stillinger1993,Jagla2001,Franzese2001,Gibson2006,Skibinsky2004}. This effectively creates an additional zone of attraction between polymer chains, which is a common attribute that produces LLPTs in soft-core potentials \cite{Jagla2001,Franzese2001}. In these models, the atoms which penetrate the soft-core, can be regarded as bonded, which generate an additional ``effective'' attractive well due to the fact that such ``bonded'' atoms have more neighbors in their attractive range \cite{Buldyrev_Hydrophobic_2010}. However, the explicit shortening of the covalent bonds between atoms would require the development of a microscopic Hamiltonian for this phenomenon, which would be most desirable for a future study. Therefore, in this work, for simplicity, instead of shortening the length of the covalent bonds, this effect is accounted for in the model through the additional ``effective'' square-well attraction (iii). Without this potential, with characteristic energy $\epsilon_{22}$, and consequently, in the absence of polymerized atoms, no LLPT will occur. We note that this simplification is in the spirit of common semi-phenomenological models of non-ideal binary mixtures, such as the Flory-Huggins theory of polymer solutions \cite{Flory_Polymer_1941,Huggins_Solutions_1941,Rey_PIPS_1996,Luo_PIPS_2006} or a regular-solution model \cite{Hildebrand_Regular_1962}.

\begin{figure}[t]
\centering
{
\includegraphics[width=0.9\linewidth]{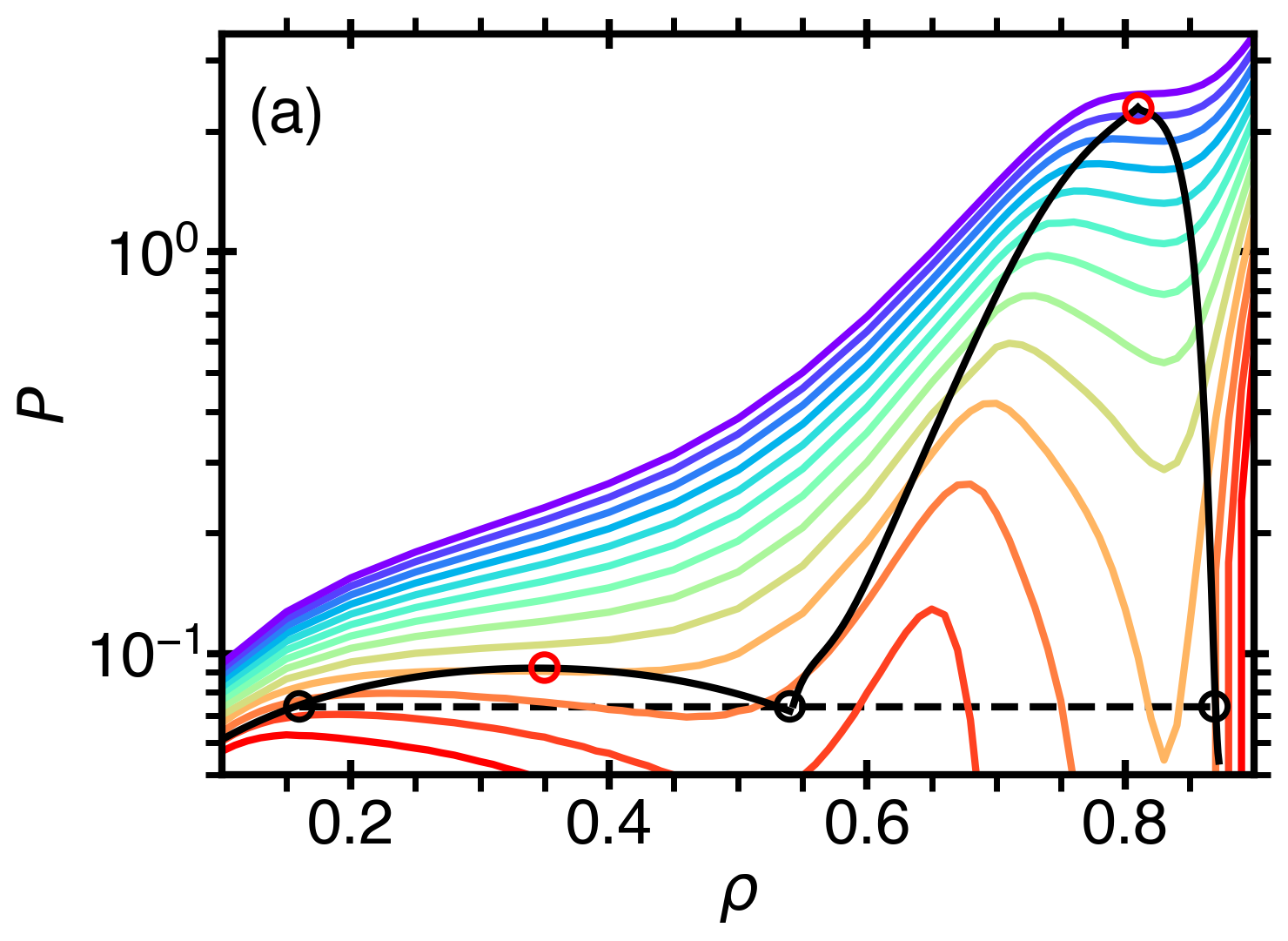}
\includegraphics[width=0.9\linewidth]{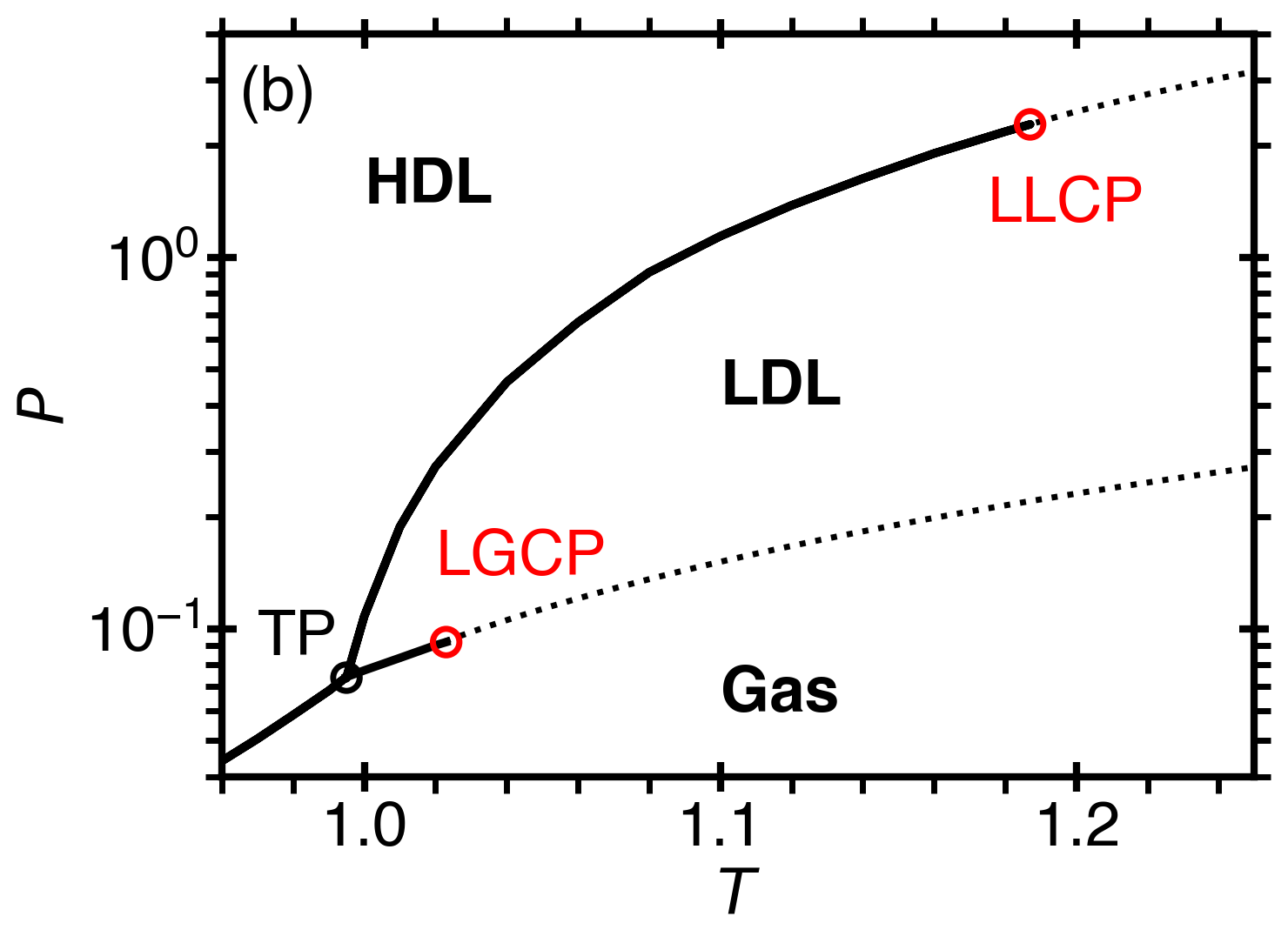}
}
\caption{Phase diagrams for the maximum-valence model (with $\epsilon_{22}=0.5$ and $\epsilon_b=1.0$) obtained in an NVT ensemble after $t=10^6$ time units. (a) The isotherms in the $P$-$\rho$ plane are $T=0.96-1.20$ (red-purple) in steps $\Delta T=0.02$. (b) The liquid-gas and liquid-liquid critical isochores in the $P$-$T$ plane are $\rho_\text{c}^\text{LG}=0.35$ and $\rho_\text{c}^\text{LL} = 0.81$ as indicated by the lower and upper dashed lines, respectively. In both figures, the liquid-gas and liquid-liquid coexistence curves are calculated via the Maxwell construction and indicated by the solid curves. The liquid-gas ($T_\text{c}^\text{LG}=1.023$, $P_\text{c}^\text{LG}=0.0922$, $\rho_\text{c}^\text{LG}=0.35$) and liquid-liquid ($T_\text{c}^\text{LL}=1.187$, $P_\text{c}^\text{LL}=2.28$, $\rho_\text{c}^\text{LL}=0.81$) critical points are indicated by the red open circles, while the triple point ($P^\text{TP}=0.0738$, $T^\text{TP}=0.995$) is indicated by the black open circles.}
\label{fig2}
\end{figure}

To verify our hypothesis, we implement these three ingredients of interactions via an event-driven MD technique \cite{Alder1959,Rapaport2004}; in particular, we use a discrete MD package (DMD) that only includes particles interacting through spherically-symmetric step-wise potentials, which may form bonds via reversible reactions \cite{Buldyrev_Application_2008}. We simulate an NVT ensemble of $N=1000$ atoms in a cubic box with periodic boundaries at various constant densities and temperatures. The temperature is controlled by a Berendsen thermostat \cite{Berendsen1984}. The van der Waals and covalent-bonding interactions are implemented by separating each atom into two overlapping hard spheres (a core and a shell), with the same diameter $\sigma$ and mass $m$, see Figs.~\ref{fig1}(b-d). The connection between the core and its shell is represented by an infinite square-well potential of width $d\ll\sigma$. The cores and shells of different atoms do not interact with each other. The core represents the atom without its valence electrons. It interacts with other cores via a wide potential well with depth $\epsilon$ and width $w = 0.4\sigma$ (the parameters are chosen as an example, Fig.~\ref{fig1}b), which models the van der Waals interactions in the system. Meanwhile, the shell represents the outer valence electron cloud. It interacts with other shells via a narrow potential well with depth $\epsilon_b = \epsilon$ and width $w_b=0.02\sigma$ (Fig.~\ref{fig1}c), which models the breaking and forming of covalent bonds. In the absence of the shell, this system has a liquid-gas critical point (LGCP) at $\rho_\text{c}^\text{LG}=N/V=0.35\pm0.05$, $T_\text{c}^\text{LG}=1.04\pm0.01$, and $P_\text{c}^\text{LG}=0.094\pm0.005$ \cite{Skibinsky2004}, well above the equilibrium crystallization line, which we force to be at low temperature by selecting the appropriate width, $w$, of the potential. We note that all physical parameters are normalized by the appropriate combination of mass $m$, length $\sigma$, and energy $\epsilon$ units, as used in Ref. \cite{Skibinsky2004}. When the shell interactions are included and the system may form covalent bonds, the location of the LGCP changes, but not significantly. In addition to the wide and narrow wells, we introduce an additional attractive potential well (with depth $\epsilon_{22}=0.5\epsilon$ and width $w_{22}=0.3\sigma$, Fig.~\ref{fig1}d) for the van der Waals interaction between the shells of the atoms with two bonds (both in the state S$_2$), which are not chemically bonded to each other.

\begin{figure}[t]
	\centering
{
\includegraphics[width=\linewidth]{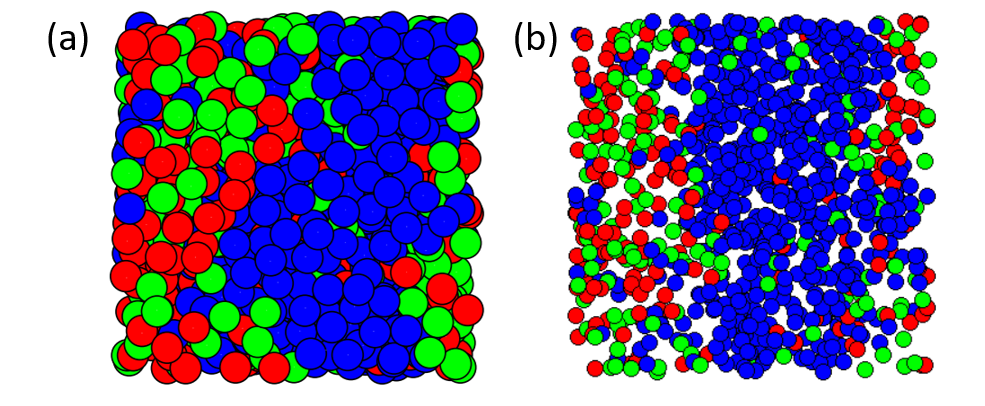}
}
\caption{Simulation snapshots of the system exhibiting phase segregation at $T=1.00$ and $\rho=0.75$ in the LL coexistence region for a) $N = 1000$ and b) $N = 8000$ (in which the image size is reduced by a factor of two). Red, green, and blue spheres indicate atom states: S$_0$, S$_1$ and  S$_2$, respectively.}
\label{fig6}
\end{figure}

We note that during either the formation or breaking of a bond, the new state of the reacting particles may modify the potential energy of the interactions with their non-bonded neighboring particles \cite{Buldyrev_Application_2008}. In our model, this occurs when particles in the state S$_1$ convert to the state S$_2$ (or vice versa). To maintain the conservation of energy, we calculate the change of the total potential energy, $\Delta U$, due to the change of the state of the reacting particles and subtract it from the kinetic energy of the reacting pair. As a consequence, the equations for computing the new velocities \cite{Buldyrev_Application_2008} may not have real solutions. In this case, the bond will not form or break, and the reacting particles will conserve their states through an elastic collision.

\begin{figure*}[ht!]
\centering
{
\includegraphics[width=0.32\linewidth]{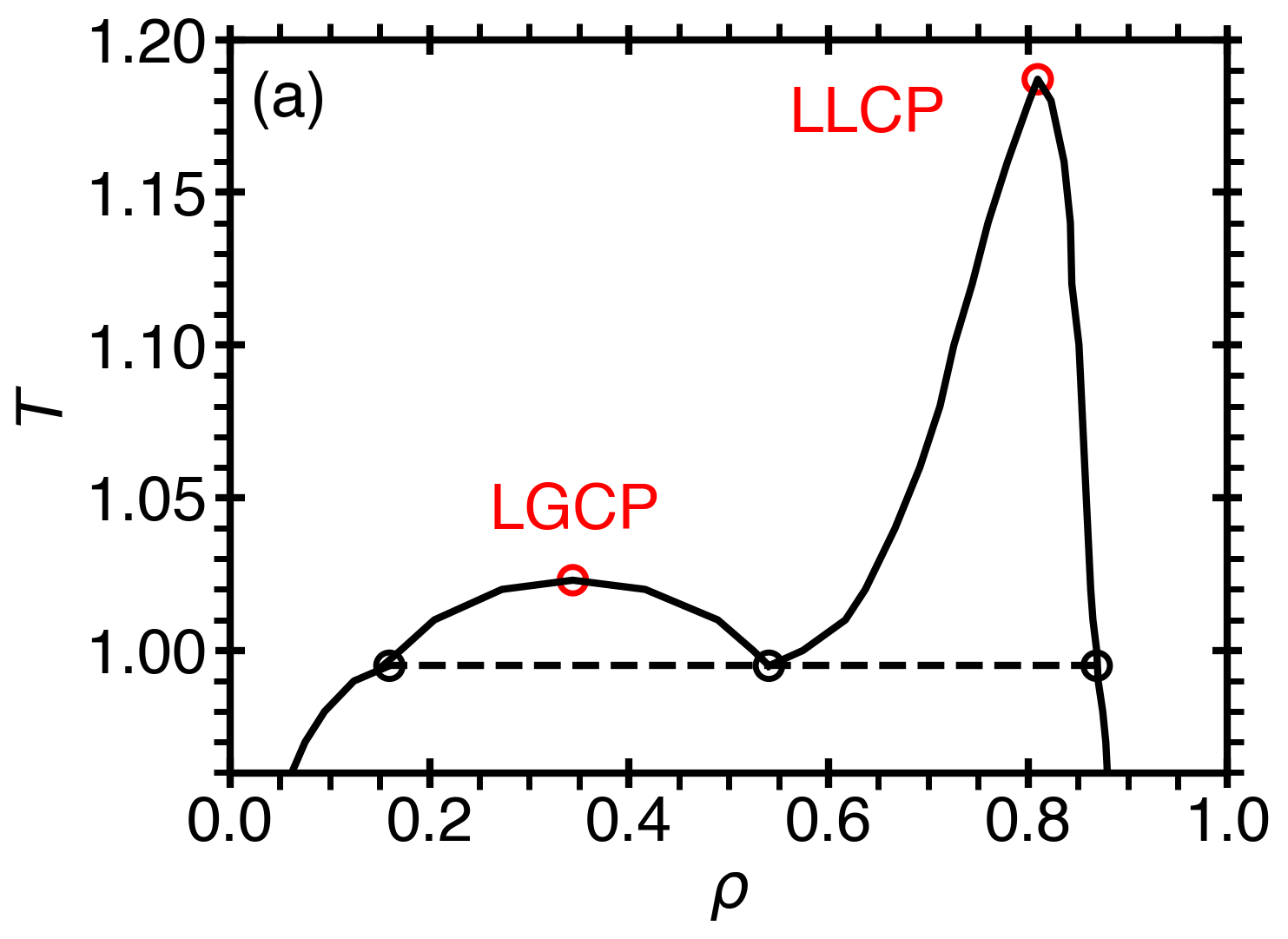}
\includegraphics[width=0.32\linewidth]{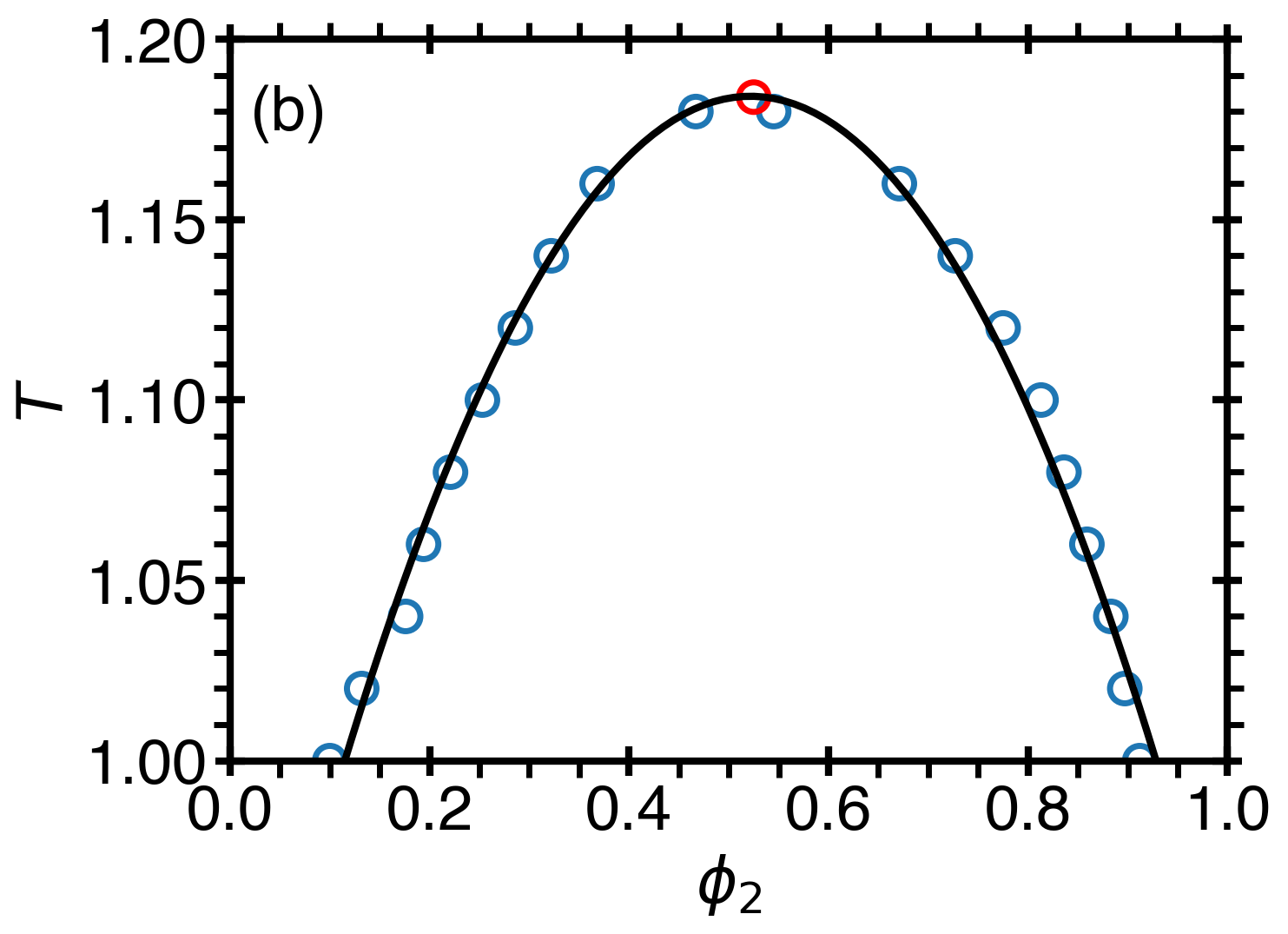}
\includegraphics[width=0.32\linewidth]{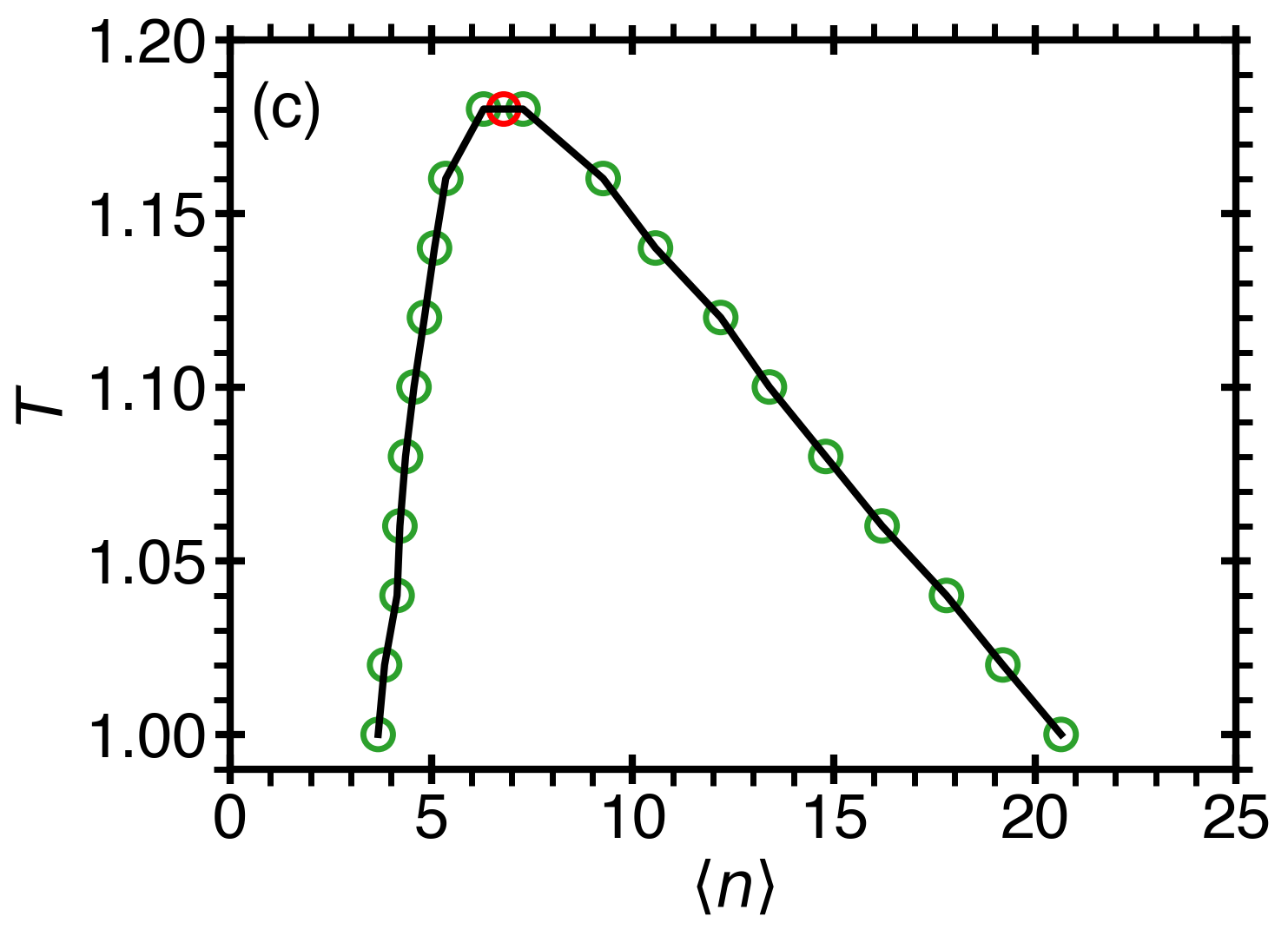}
}
\caption{(a) $T$-$\rho$ phase diagram for the maximum-valence model (with $\epsilon_{22}=0.5$ and $\epsilon_b=1.0$) obtained in an NVT ensemble after $t=10^6$ time units. The temperature dependence of the fraction of atoms with two bonds, $\phi_2$, (b) and the average chain-length, $\langle n\rangle$, (c) in two coexisting liquid phases. The simulation data in (b) is fit to a second order polynomial, while in (c) the curve is provided as a guide.}
\label{fig3}
\end{figure*}

In this work, we obtain a detailed phase diagram of the model using the values of the square-well depths and widths illustrated in Figs.~\ref{fig1}(b-d). In addition, we investigate the effect of $\epsilon_b$ and $\epsilon_{22}$ on the position of the liquid-liquid (LLCP) and liquid-gas (LGCP) critical points.

\section{Results: Liquid-Liquid Phase Transition} Figure~\ref{fig2}a illustrates isotherms on a pressure-density ($P$-$\rho$) plane, which exhibit two sets of van der Waals loops. The loops correspond to the LGCP, located at low density and pressure, and the LLCP, located at a higher density and pressure. Fig.~\ref{fig2}b illustrates the LG and LL coexistence on a $P$-$T$ plane along with the critical isochores. At the triple point (TP), the gaseous, LDL, and HDL phases coexist. In contrast to the ST2 model for water \cite{Poole1992}, but in agreement with spherically symmetric models \cite{Franzese2001,Luo2015}, the  $P$-$T$ line of the LL coexistence has a positive slope. Simulation snapshots depicted in Fig.~\ref{fig6} show the segregation of polymer-rich, HDL, and polymer-poor, LDL, phases. 

Figure~\ref{fig3}a presents the LG and LL coexistence curves on a $T$-$\rho$ phase diagram. Although there is a distribution of polymer chains with varying lengths, a simple way to characterize the degree of polymerization is to find the fractions $\phi_0$, $\phi_1$ and $\phi_2$ of atoms in states S$_0$, S$_1$ and S$_2$. Due to the conservation of the number of atoms, $\phi_0+\phi_1+\phi_2 = 1$. The fraction $\phi_2$ was computed based on the asymmetric LL coexistence curve (Fig.~\ref{fig3}a). Remarkably, $\phi_2$ was found to be symmetric and centered around $\phi_2=0.5$ as shown in Fig.~\ref{fig3}b. Consequently, the sum $\phi_0 + \phi_1 = 1-\phi_2$ has the same symmetry. This feature suggests that $1-\phi_2$ may be viewed as the appropriate order parameter for the LLPT coupled with polymerization. In contrast, the density, $\rho -\rho_\text{c}^\text{LG}$, is the order parameter for the LGPT, as commonly accepted. The symmetric nature of $\phi_2$, and the fact that S$_1$ atoms are the intermediate states in the formation of polymer chains, enables a two-state thermodynamic approach \cite{Anisimov2018} by reducing this model to two alternative states, with fractions $\phi_2$ and $\phi_0$ + $\phi_1$.

The LLPT coexistence curve on the $T$-$\langle n\rangle$ plane (Fig.~\ref{fig3}c), where $\langle n\rangle$ is the average length of a polymer chain among those containing at least one atom in state S$_2$, namely trimers or longer polymer chains. The strong temperature dependence of $\langle n\rangle$ in the phase segregation region proves that the LLPT is associated with polymerization. Neither $\phi_2$ nor $\langle n\rangle$ shows any discontinuity as a function of density and temperature, although $\langle n\rangle$ shows a strong asymmetry toward the HDL phase.

\section{Location of the Critical Points} 
\begin{figure}[t]
    \centering
    \includegraphics[width=0.49\linewidth]{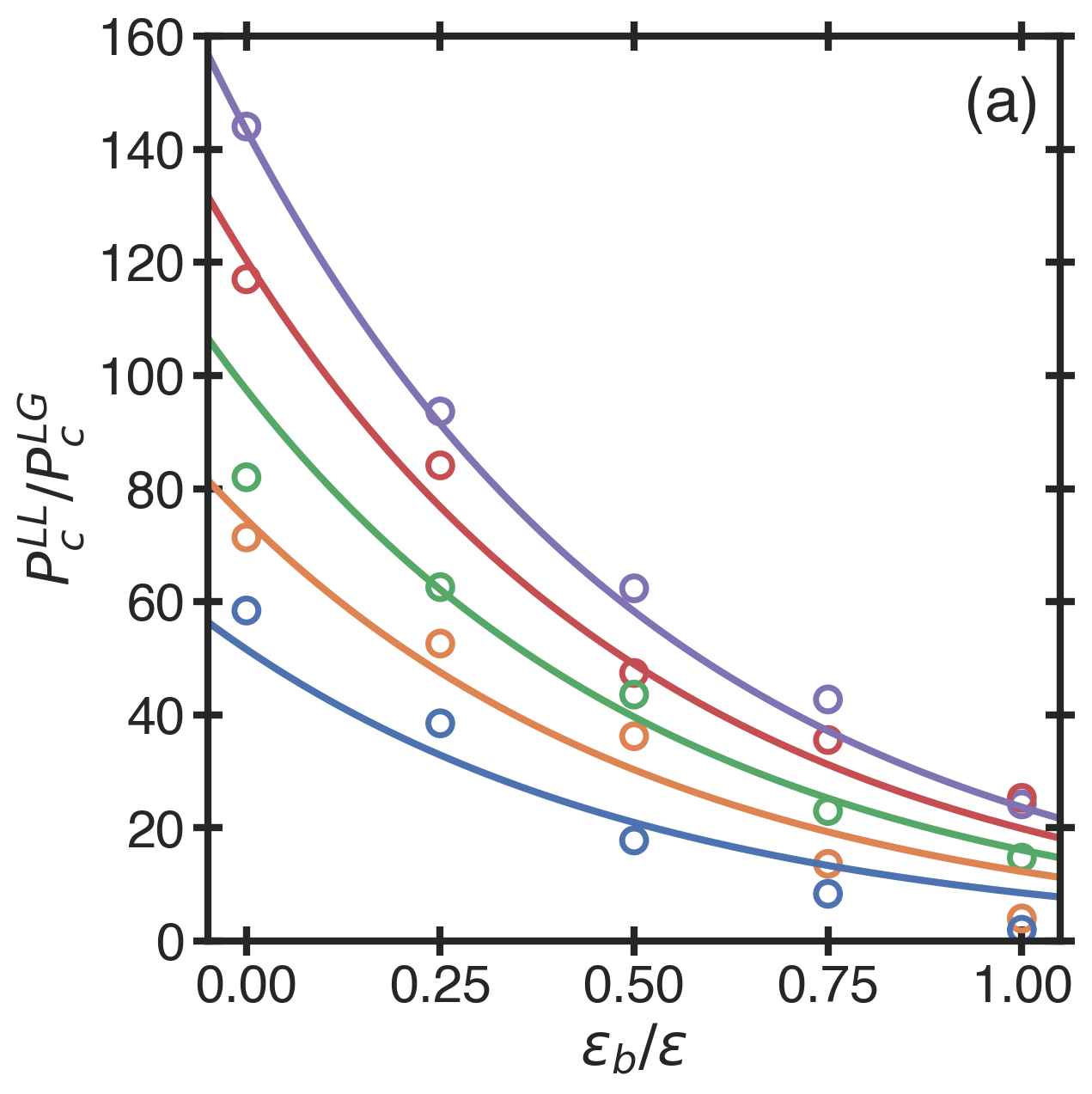}
    \includegraphics[width=0.49\linewidth]{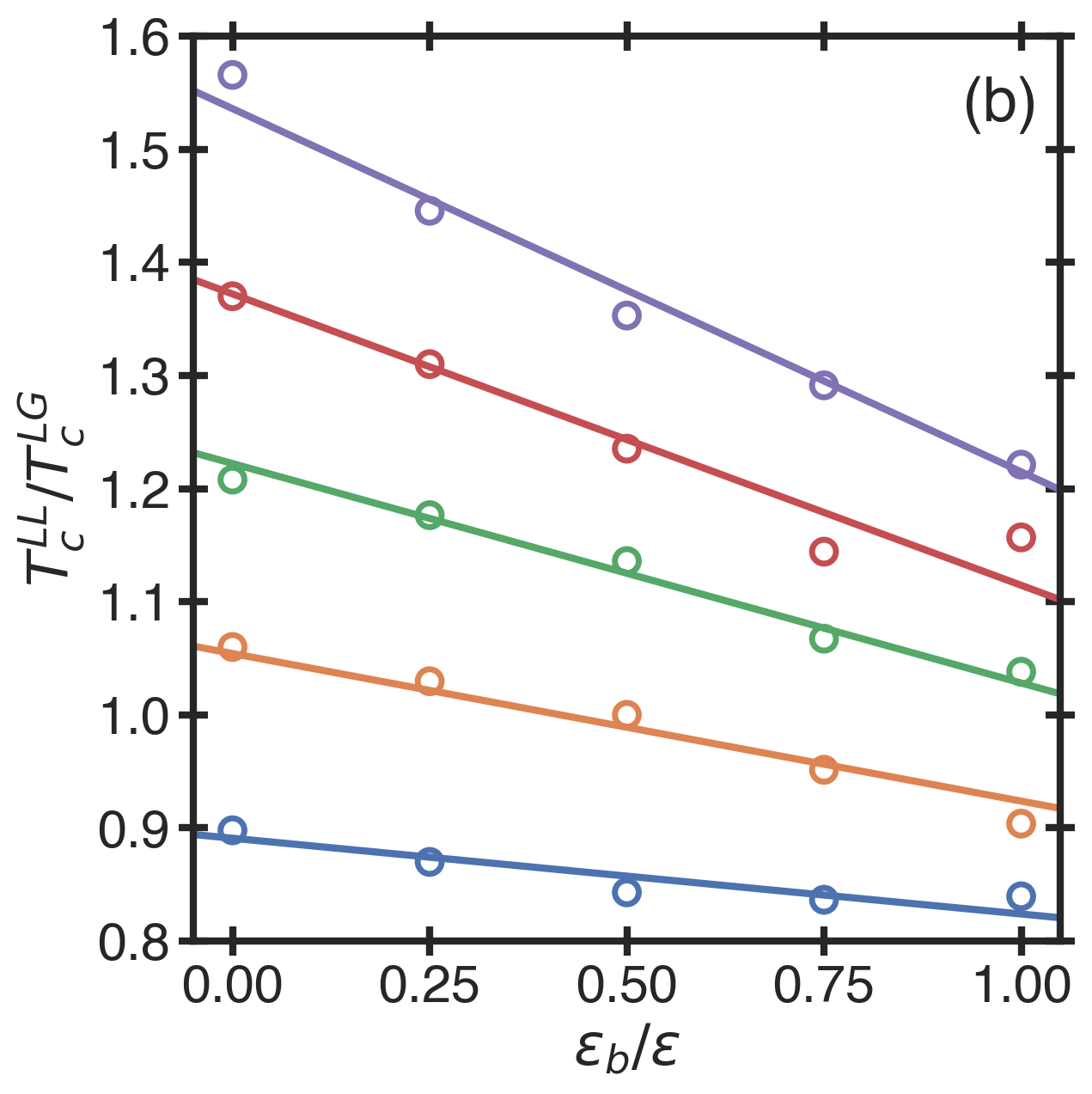}
    \includegraphics[width=0.49\linewidth]{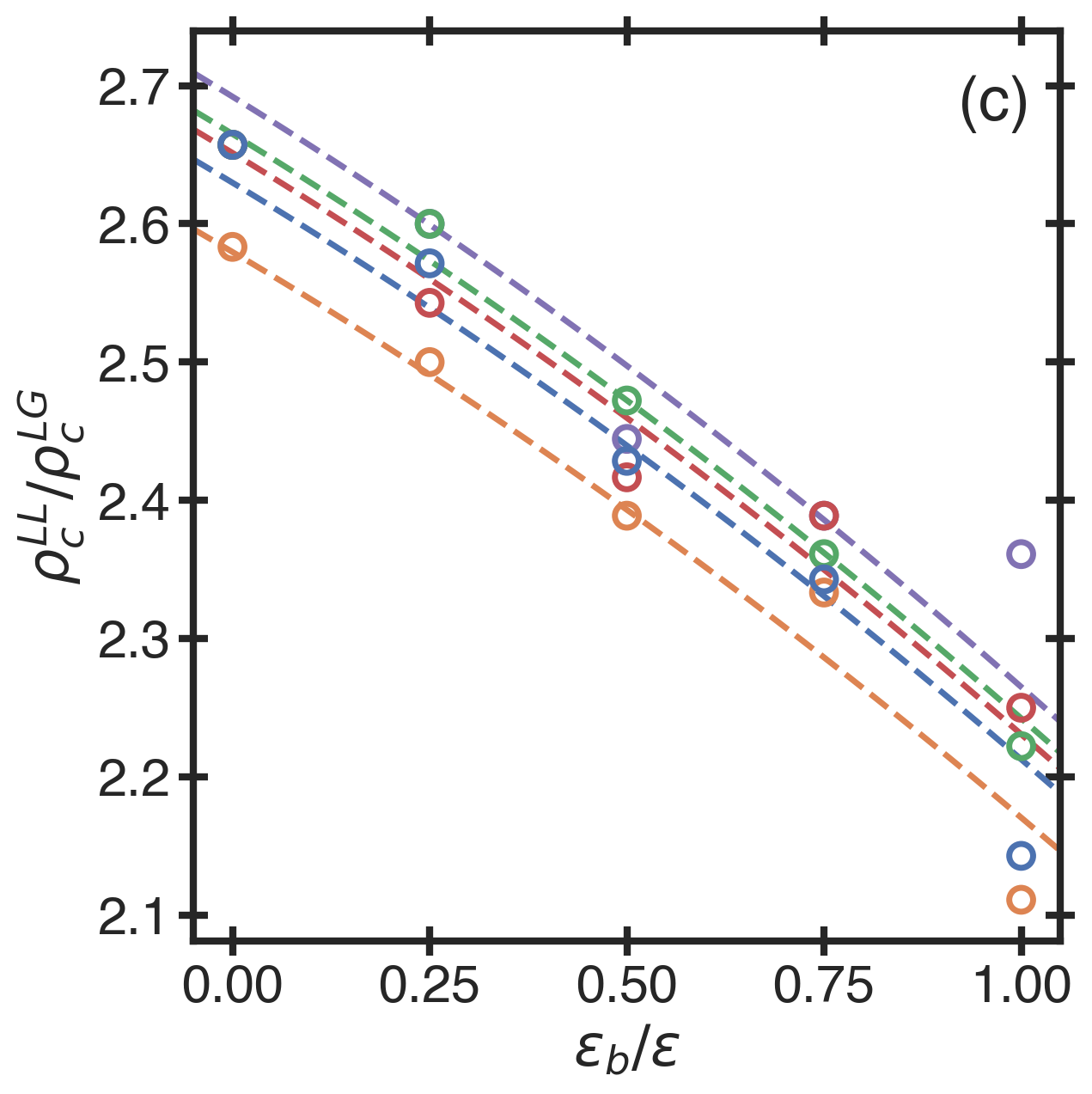}
    \includegraphics[width=0.49\linewidth]{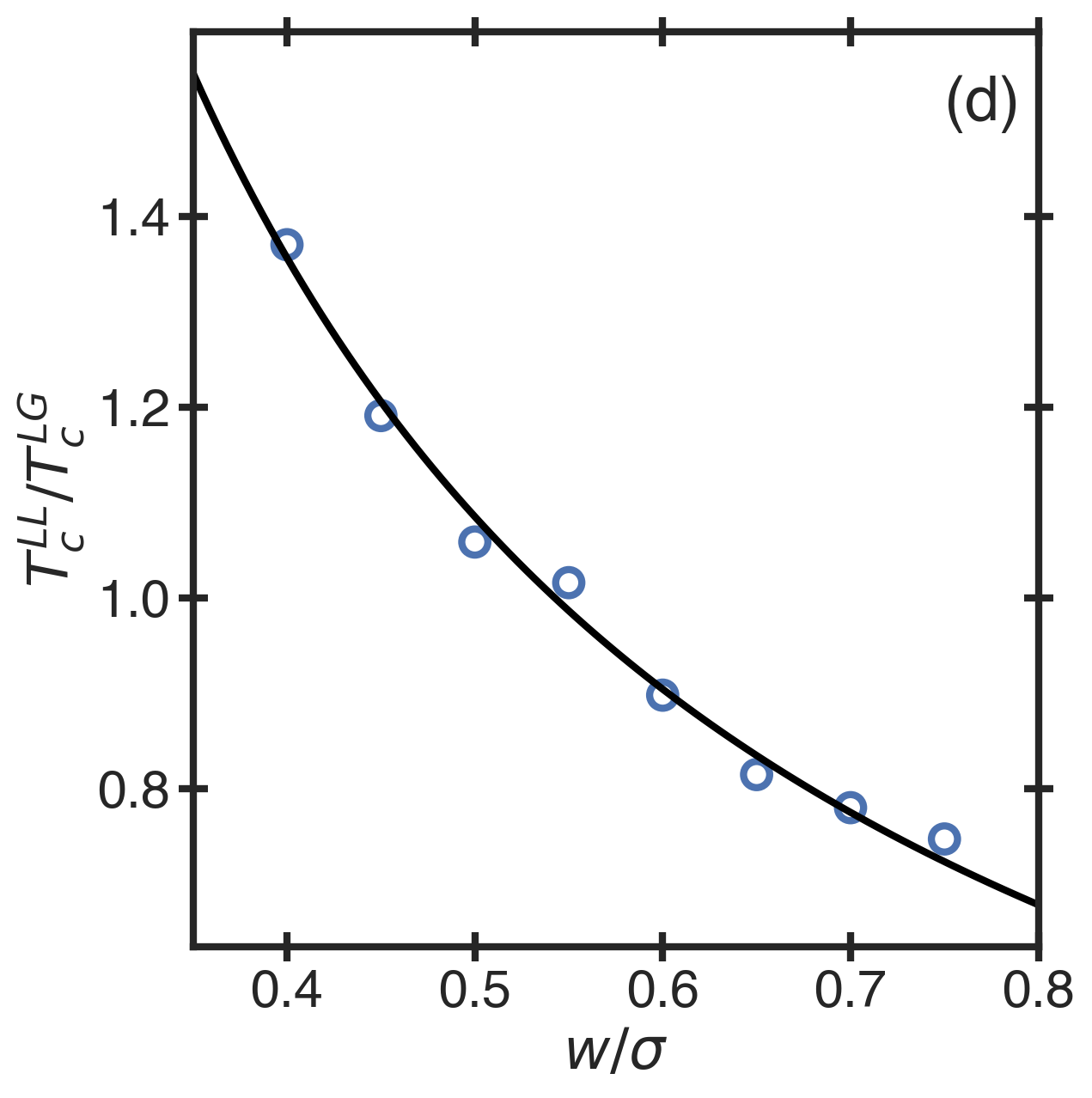}
    \caption{{The effect of different interaction parameters on the critical point locations in the maximum-valence model. In (a-c), for $w=0.4\sigma$ for increasing the bond energy, $\epsilon_b$, with interaction energy between bonded atoms: $\epsilon_{22}=0.55\epsilon$ (purple), $\epsilon_{22}=0.50\epsilon$ (red), $\epsilon_{22}=0.45\epsilon$ (green), $\epsilon_{22}=0.4\epsilon$ (orange), and $\epsilon_{22}=0.35\epsilon$ (blue). In (a), the ratio of the critical pressures exponentially decreases as $P_\text{c}^\text{LL}/P_\text{c}^\text{LG}\sim 460(\epsilon_{22}⁄\epsilon) \exp\left[-\epsilon_b/(0.55\epsilon)\right]$, while in (b) the ratio of the critical temperatures decreases linearly as $T_\text{c}^\text{LL}/T_\text{c}^\text{LG}\sim -1.27\epsilon_b \epsilon_{22}/\epsilon^2$, and in (c), the ratio of critical densities shows a general decreasing trend indicated by the second-order polynomial guidelines (dashed). In (d), the ratio of the critical temperatures is inversely related to the width of the van der Waals interaction well, $T_\text{c}^\text{LL}/T_\text{c}^\text{LG}=0.54\sigma/w$, for $\epsilon_{22}=0.5\epsilon$ and $\epsilon_b=0.5\epsilon$.} }
    \label{Fig_CP_Locs}
\end{figure}

With the values of parameters considered in the previous sections, the system acquires a LLPT terminating at a second critical point located at $T_\text{c}^\text{LL}=1.187$, $P_\text{c}^\text{LL}=2.28$ and $\rho_\text{c}^\text{LL}=0.81$. We have investigated the dependency of the location of the LLCP and LGCP on the the three key parameters of the model: the range of van der Waals interactions $w$, the bond strength $\epsilon_b$, and the interaction energy between bonded and unbonded atoms, $\epsilon_{22}$, {presented in Fig.~\ref{Fig_CP_Locs}. Each of the parameter sets produced a LLCP at much higher pressures than the LGCP, which practically remains the same as the square-well model without bonds \cite{Xu_LLPT_2005}. We observed that the reduction of the interaction energy, $\epsilon_{22}$ proportionally reduces $T_\text{c}^\text{LL}$ and $P_\text{c}^\text{LL}$ (see Figs.~\ref{Fig_CP_Locs}a,b). This indicates that the attraction between atoms in state S$_2$ is crucial for the existence of the LLPT, since decreasing the interaction energy further decreases $P_\text{c}^\text{LL}$ to negative pressures and, eventually, to a point in the metastable region below the liquid-gas coexistence curve, where the LLCP effectively disappears.} We find that for all simulations with $\epsilon_{22}>0.55\epsilon$, the LGCP moves into the metastable region of the LLCP and effectively disappears{, while} for $\epsilon_{22}<0.35\epsilon$, the interactions between polymer chains are too weak to produce the LLCP. 

{In contrast, increasing the bond strength, $\epsilon_b$, does not indicate that the LLCP is going to disappear. Increasing the bond energy produces a slight increase in $T_\text{c}^\text{LL}$ and $\rho_\text{c}^\text{LL}$, while producing a significant decrease in $P_\text{c}^\text{LL}$ as shown in Figs.~\ref{Fig_CP_Locs}(a-c).} We note that when the bond energy becomes larger than the van der Waals interaction energy, $\epsilon_b>\epsilon$, the LLCP drops to negative pressures (or in some cases, drops below the crystallization line) and may disappear.  This indicates that increasing the strength of polymer bonds is not crucial for the existence of the LLPT. {In addition, we found that by increasing the width of the van der Waals interaction potential attraction between cores, $w$, causes $T_\text{c}^\text{LL}$ to increase slightly, while producing a greater increase in $T_\text{c}^\text{LG}$ (see Fig.~\ref{Fig_CP_Locs}d). Meanwhile, increasing $w$ causes the inverse effect in $P_\text{c}^\text{LL}$ and $P_\text{c}^\text{LG}$, while leaving $\rho_\text{c}^\text{LL}$ and $\rho_\text{c}^\text{LG}$ essentially unaffected. We note that the maximum-valence model produces a LLCP even when the bond energy is zero (as shown for in Figs.~\ref{Fig_CP_Locs}(a-c) for $w=0.4\sigma$). This indicates that the interplay between the van der Waals interaction and the attraction between bonded atoms is crucial to generate the LLPT, while the strength of the bond is secondary to this effect.}

{To compare the critical point behavior in the maximum-valence model with sulfur, we considered the ratio of the LLCP and LGCP critical-point parameters. For sulfur, these ratios are $P_\text{c}^\text{LL}/P_\text{c}^\text{LG}=104$, $T_\text{c}^\text{LL}/T_\text{c}^\text{LG}=0.78$, and $\rho_\text{c}^\text{LL}/\rho_\text{c}^\text{LG}=3.4$ \cite{Henry2020,Lide2003}. Fig.~\ref{Fig_CP_Locs} depicts the behavior of the critical-point parameters for the maximum valence model.} As illustrated in Fig.~\ref{Fig_CP_Locs}a, we find that the ratio of critical pressures scales linearly with $\epsilon_{22}$ and exponentially with $\epsilon_b$ as $P_\text{c}^\text{LL}/P_\text{c}^\text{LG}\sim 460(\epsilon_{22}/\epsilon)e^{-\epsilon_b/(0.55\epsilon)}$. We also find that the liquid-liquid critical temperature is linearly related to $\epsilon_b$ and $\epsilon_{22}$ as $T_\text{c}^\text{LL}/T_\text{c}^\text{LG}\sim -1.27\epsilon_b\epsilon_{22}/\epsilon^2$ {(see Fig.~\ref{Fig_CP_Locs}b)}, {while also being} inversely proportional to $w$ as the ratio $T_\text{c}^\text{LL}/T_\text{c}^\text{LG}\approx 0.54\sigma/w$ {(see Fig.~\ref{Fig_CP_Locs}d). Meanwhile, as illustrated in Fig.~\ref{Fig_CP_Locs}c}, the critical density shows a general decreasing trend with increase of $\epsilon_b$. {From the general trends presented in Fig.~\ref{Fig_CP_Locs}, we find that the best parameter sets in the maximum-valence model that produce ratios that match with sulfur exist for large van der Waals interaction potential, $w$, and small bond energies, $\epsilon_b\ll\epsilon$ at large interaction potentials between bonded atoms $\epsilon_{22}$. For $w=0.7\sigma$, $\epsilon_{22} = 0.5\epsilon$, and $\epsilon_b=0.0\epsilon$, producing $P_\text{c}^\text{LL}/P_\text{c}^\text{LG}=69.5$, $T_\text{c}^\text{LL}/T_\text{c}^\text{LG}=0.78$, and $\rho_\text{c}^\text{LL}/\rho_\text{c}^\text{LG}=3.30$, values which are close to the ratio in sulfur \cite{Henry2020,Lide2003}}.

\section{Further Comparisons with the Behavior of Sulfur}
Qualitatively, the phase diagram of sulfur matches that of the maximum-valence model with a specific set of interaction parameters. In sulfur, the LGCP is located at $T_\text{c}^\text{LG}=\SI{1314}{\kelvin}$, $P_\text{c}^\text{LG}=\SI{20.7}{\mega\pascal}$, and $\rho_\text{c}^\text{LG}=\SI{563}{\kilogram/\meter^3}$ \cite{Lide2003}, while the LLCP is located at $T_\text{c}^\text{LL}=\SI{1035}{\kelvin}$, $P_\text{c}^\text{LL}=\SI{2.15}{\giga\pascal}$, and $\rho_\text{c}^\text{LL}\approx \SI{2000}{\kilogram/\meter^3}$ \cite{Henry2020}, such that the ratio of the LL to LG critical parameters qualitatively matches the predictions of the maximum-valence model. {We note that the behavior of sulfur is more complicated away from the LLPT since liquid sulfur contains octamers that (above the lambda transition \cite{Tobolsky_Sulfur_1959,Sauer_Lambda_1967,Bellissent_Sulfur_1994,Kozhevnikov_Sulfur_2004,Tobolsky_Selenium_1960}) are to be broken down upon heating before polymerization can occur \cite{Tobolsky_Sulfur_1959}. Since in the considered formulation of the maximum-valence model, we consider atoms that form linear polymers, this mimics the valence structure and bond formation of sulfur in the vicinity of the LLPT.}

\begin{figure}[t]
    \centering
    \includegraphics[width=\linewidth]{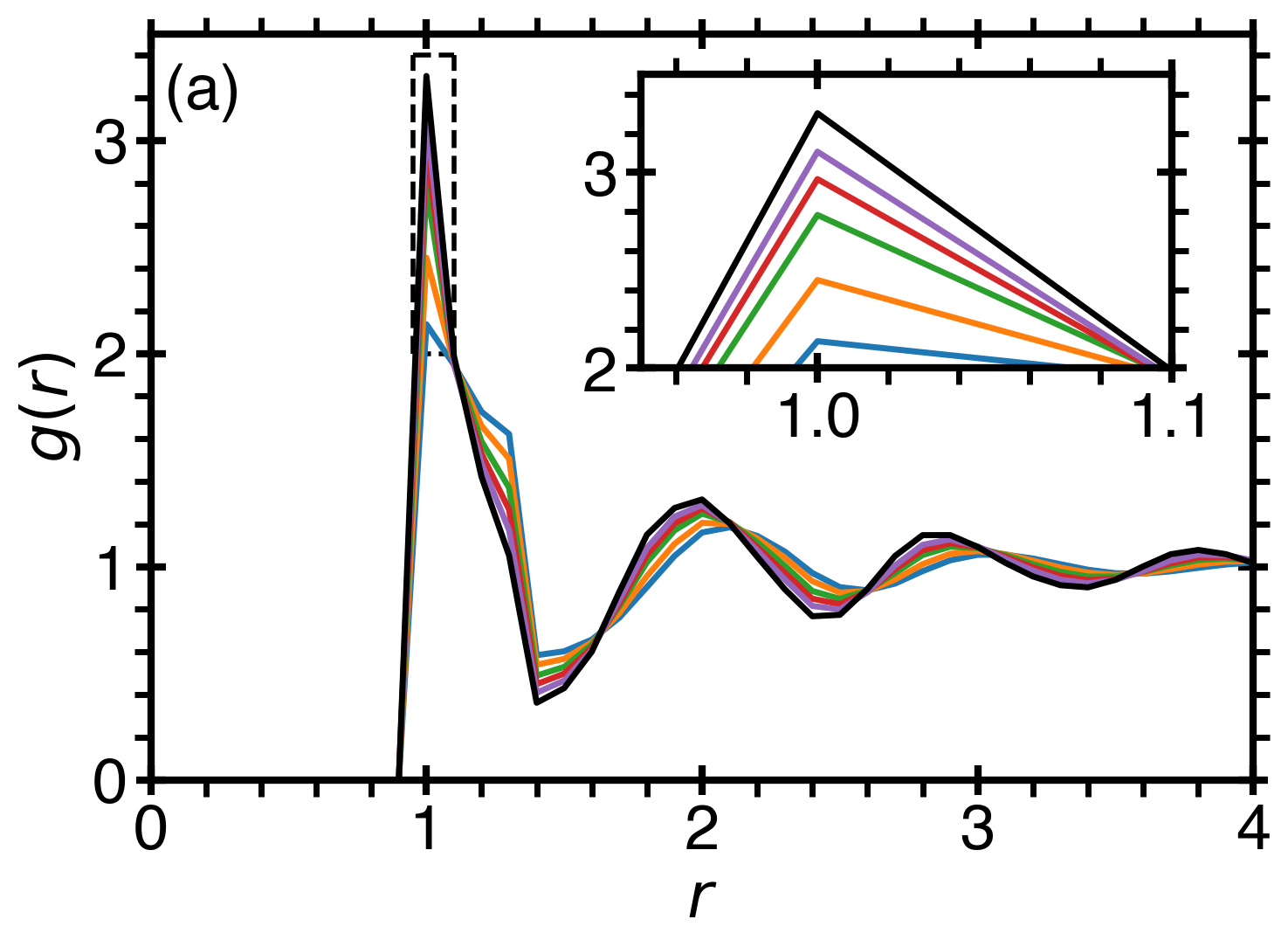}
    \includegraphics[width=\linewidth]{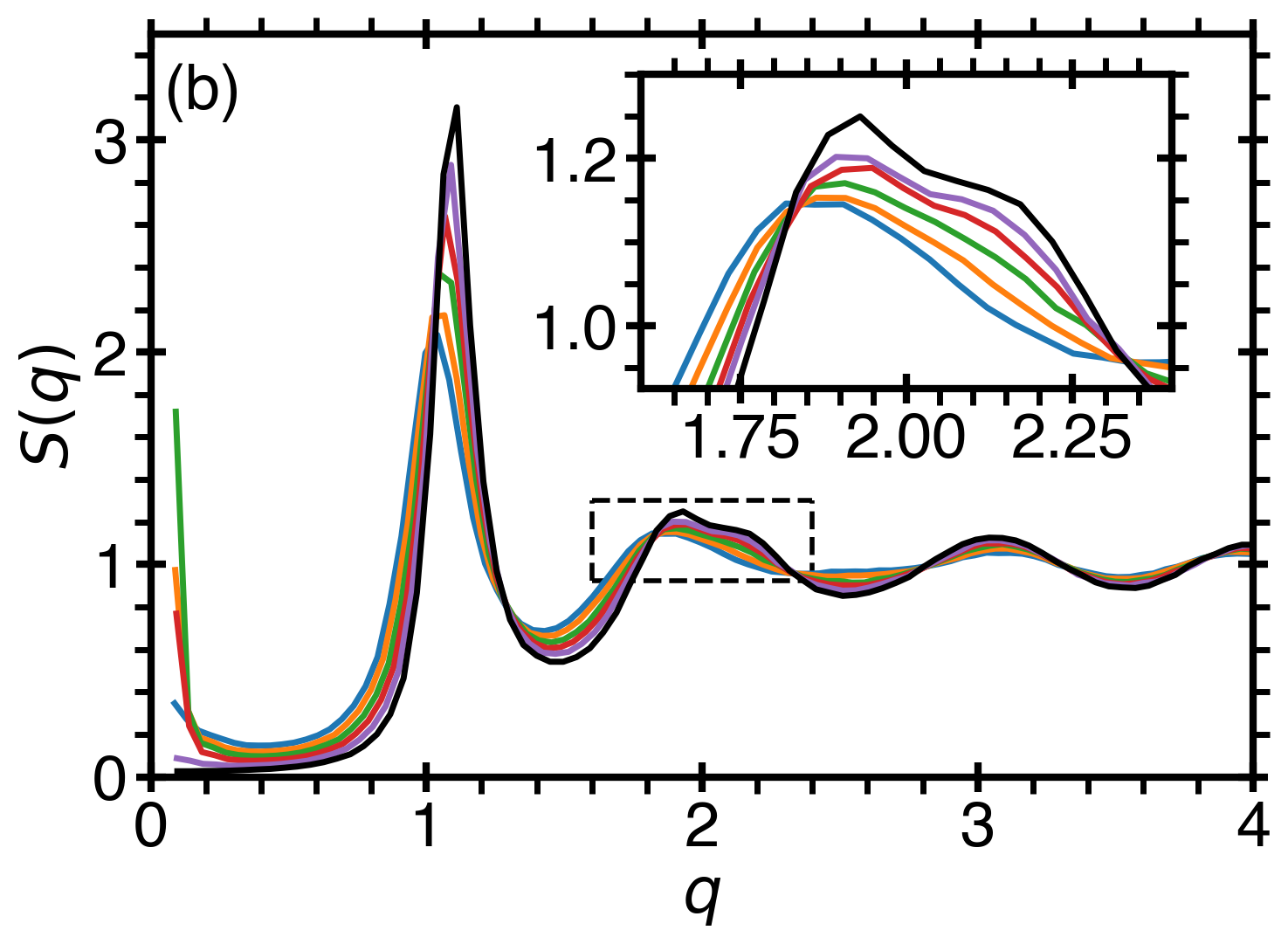}
    \caption{{(a) The density correlation function $g(r)$ and (b) the structure factor $S(q)$ across the liquid-liquid transition at $T=1.00$ for densities of $\rho=0.65$ (blue), $\rho=0.70$ (orange), $\rho=0.75$ (green), $\rho=0.80$ (red), $\rho=0.85$ (purple), and $\rho=0.90$ (black). In (a), the sharp peak, around $r=1$ (in units of $\sigma$), corresponds to the length of the covalent bond, which increases upon increasing density. Simultaneously, in (b), the maximum of the structure factor (the first peak) shifts to larger wavenumbers upon increasing density, while the second peak acquires a characteristic bump, similar to what was recently observed in sulfur \cite{Henry2020}. The divergence of the structure factor at $q=0$ indicates the divergence of the isothermal compressibility in the vicinity of the LLCP. The insets (dashed boxes) highlight the behavior of the maximum of the correlation function and second peak of the structure factor.} }
    \label{Fig_Struct_Fact}
\end{figure}

Also, the computed structure factor contains qualitative similarities with the LLPT in sulfur. {In Fig.~\ref{Fig_Struct_Fact}, we depict the structural differences between the LDL and HDL phases through the density correlation function, $g(r)$, and the structure factor, $S(q)$, for several densities at constant temperature near the liquid-liquid coexistence (computed for the atom cores). In Fig.~\ref{Fig_Struct_Fact}a, the $g(r)$ shows a sharp peak corresponding to the covalent bond length $r=1.02\sigma$, in the HDL phase. Correspondingly, the structure factor shows a shift in the first peak to a larger wavenumber $q$, while the second peak changes due to polymerization. This change is similar to what was observed in a recent experiment on sulfur \cite{Henry2020}. In addition, $S(q)$ shows a dramatic increase as $q\to 0$ for the points corresponding to the equilibrium between two liquid phases (see Fig.~\ref{Fig_Struct_Fact}b), which is indicative of the divergence of the isothermal compressibility.} We note that in this work, we find a gas-LDL-HDL triple point, while in the recent experimental work on sulfur \cite{Henry2020}, the solid-LDL-HDL triple point is observed. In principle, this triple point may be reproduced in the maximum-valence model by fine-tuning the parameters, which requires further investigation.

\section{Conclusion} 
The maximum-valence model describes liquid polyamorphism in a variety of chemically-reacting fluids. By tuning the maximum valency, $z$ (maximum coordination number), of the model, the liquid-liquid phase transitions (LLPT) in these systems could be investigated. We show that when the bonded atoms attract each other stronger than the non-bonded atoms or when the bonded and non-bonded atoms repel each other, the LLPT is generated by the coupling between phase separation and the chemical reaction. In this work, we compared the model with $z=2$ to the behavior of liquid sulfur. Our results show that the LLPT predicted by the model qualitatively reproduces the LLPT in sulfur at a high pressure and temperature. 

The model could also be used to study the LLPTs in systems with other maximum valence numbers. For instance, when $z=1$, the LLPT is induced by dimerization (e.g. hydrogen at extremely large pressures {\cite{Ohta_H_2015,McWilliams_H_2016,Zaghoo_H_2013,Zaghoo_H_2016,Zaghoo_H_2017}}). For $z\ge 3$, the LLPT could be induced either by gelation or by molecular network formation \cite{Zaccarelli2005}. For example, it could be used to model the phase behavior of liquid phosphorous with $z=3$ \cite{Katayama2000,Katayama_Phos2_2004} {as well as silicon \cite{Sciorino_Silicon_2011}, silica \cite{Lascaris_Silica_2014,Chen_Silica_2017}, or} supercooled water with $z>3$ \cite{Gallo2016,Debenedetti2}. In a future study, the two-state thermodynamics of liquid polyamorphism \cite{Anisimov2018, Caupin2021,Longo2021} could be applied to these systems to develop the equation of state, which would determine the anomalies of the physical properties in this system, especially near the critical points.

\begin{acknowledgments}
The authors thank Fr\'ed\'eric Caupin, Pablo G. Debenedetti, Francesco Sciortino, and Eugene I. Shakhnovich for useful discussions. This work is a part of the research collaboration between the University of Maryland, Princeton University, Boston University, and Arizona State University supported by the National Science Foundation. The research at Boston University was supported by NSF Award No. 1856496, while the research at the University of Maryland was supported by NSF award no. 1856479. S.V.B. acknowledges the partial support of this research through Bernard W. Gamson Computational Science Center at Yeshiva College.
\end{acknowledgments}

\bibliography{refs}

\begin{thebibliography}{67}%
\makeatletter
\providecommand \@ifxundefined [1]{%
 \@ifx{#1\undefined}
}%
\providecommand \@ifnum [1]{%
 \ifnum #1\expandafter \@firstoftwo
 \else \expandafter \@secondoftwo
 \fi
}%
\providecommand \@ifx [1]{%
 \ifx #1\expandafter \@firstoftwo
 \else \expandafter \@secondoftwo
 \fi
}%
\providecommand \natexlab [1]{#1}%
\providecommand \enquote  [1]{``#1''}%
\providecommand \bibnamefont  [1]{#1}%
\providecommand \bibfnamefont [1]{#1}%
\providecommand \citenamefont [1]{#1}%
\providecommand \href@noop [0]{\@secondoftwo}%
\providecommand \href [0]{\begingroup \@sanitize@url \@href}%
\providecommand \@href[1]{\@@startlink{#1}\@@href}%
\providecommand \@@href[1]{\endgroup#1\@@endlink}%
\providecommand \@sanitize@url [0]{\catcode `\\12\catcode `\$12\catcode
  `\&12\catcode `\#12\catcode `\^12\catcode `\_12\catcode `\%12\relax}%
\providecommand \@@startlink[1]{}%
\providecommand \@@endlink[0]{}%
\providecommand \url  [0]{\begingroup\@sanitize@url \@url }%
\providecommand \@url [1]{\endgroup\@href {#1}{\urlprefix }}%
\providecommand \urlprefix  [0]{URL }%
\providecommand \Eprint [0]{\href }%
\providecommand \doibase [0]{https://doi.org/}%
\providecommand \selectlanguage [0]{\@gobble}%
\providecommand \bibinfo  [0]{\@secondoftwo}%
\providecommand \bibfield  [0]{\@secondoftwo}%
\providecommand \translation [1]{[#1]}%
\providecommand \BibitemOpen [0]{}%
\providecommand \bibitemStop [0]{}%
\providecommand \bibitemNoStop [0]{.\EOS\space}%
\providecommand \EOS [0]{\spacefactor3000\relax}%
\providecommand \BibitemShut  [1]{\csname bibitem#1\endcsname}%
\let\auto@bib@innerbib\@empty
\bibitem [{\citenamefont {Stanely}(2013)}]{Stanley_Liquid_2013}%
  \BibitemOpen
  \bibfield  {author} {\bibinfo {author} {\bibfnamefont {H.~E.}\ \bibnamefont
  {Stanely}},\ }\href@noop {} {\emph {\bibinfo {title} {Liquid
  Polymorphism}}},\ edited by\ \bibinfo {editor} {\bibfnamefont {A.~R.~D.}\
  \bibnamefont {Stuart A.~Rice}},\ \bibinfo {series} {Advances in Chemical
  Physics}, Vol.\ \bibinfo {volume} {152}\ (\bibinfo  {publisher} {JohnWiley \&
  Sons},\ \bibinfo {year} {2013})\BibitemShut {NoStop}%
\bibitem [{\citenamefont {Anisimov}\ \emph {et~al.}(2018)\citenamefont
  {Anisimov}, \citenamefont {Duška}, \citenamefont {Caupin}, \citenamefont
  {Amrhein}, \citenamefont {Rosenbaum},\ and\ \citenamefont
  {Sadus}}]{Anisimov2018}%
  \BibitemOpen
  \bibfield  {author} {\bibinfo {author} {\bibfnamefont {M.~A.}\ \bibnamefont
  {Anisimov}}, \bibinfo {author} {\bibfnamefont {M.}~\bibnamefont {Duška}},
  \bibinfo {author} {\bibfnamefont {F.}~\bibnamefont {Caupin}}, \bibinfo
  {author} {\bibfnamefont {L.~E.}\ \bibnamefont {Amrhein}}, \bibinfo {author}
  {\bibfnamefont {A.}~\bibnamefont {Rosenbaum}},\ and\ \bibinfo {author}
  {\bibfnamefont {R.~J.}\ \bibnamefont {Sadus}},\ }\bibfield  {title} {\bibinfo
  {title} {Thermodynamics of fluid polyamorphism},\ }\href
  {https://doi.org/10.1103/PhysRevX.8.011004} {\bibfield  {journal} {\bibinfo
  {journal} {Phys. Rev. X}\ }\textbf {\bibinfo {volume} {8}},\ \bibinfo {pages}
  {011004} (\bibinfo {year} {2018})}\BibitemShut {NoStop}%
\bibitem [{\citenamefont {Tanaka}(2020)}]{Tanaka_Liquid_2020}%
  \BibitemOpen
  \bibfield  {author} {\bibinfo {author} {\bibfnamefont {H.}~\bibnamefont
  {Tanaka}},\ }\bibfield  {title} {\bibinfo {title} {Liquid–liquid transition
  and polyamorphism},\ }\href
  {https://doi.org/https://doi.org/10.1063/5.0021045} {\bibfield  {journal}
  {\bibinfo  {journal} {J. Chem. Phys.}\ }\textbf {\bibinfo {volume} {153}},\
  \bibinfo {pages} {130901} (\bibinfo {year} {2020})}\BibitemShut {NoStop}%
\bibitem [{\citenamefont {Franzese}\ \emph {et~al.}(2001)\citenamefont
  {Franzese}, \citenamefont {Malescio}, \citenamefont {Skibinsky},
  \citenamefont {Buldyrev},\ and\ \citenamefont {Stanley}}]{Franzese2001}%
  \BibitemOpen
  \bibfield  {author} {\bibinfo {author} {\bibfnamefont {G.}~\bibnamefont
  {Franzese}}, \bibinfo {author} {\bibfnamefont {G.}~\bibnamefont {Malescio}},
  \bibinfo {author} {\bibfnamefont {A.}~\bibnamefont {Skibinsky}}, \bibinfo
  {author} {\bibfnamefont {S.~V.}\ \bibnamefont {Buldyrev}},\ and\ \bibinfo
  {author} {\bibfnamefont {H.~E.}\ \bibnamefont {Stanley}},\ }\bibfield
  {title} {\bibinfo {title} {Generic mechanism for generating a liquid-liquid
  phase transition},\ }\bibfield  {journal} {\bibinfo  {journal} {Nature}\
  }\textbf {\bibinfo {volume} {409}},\ \href
  {https://doi.org/https://doi.org/10.1038/35055514}
  {https://doi.org/10.1038/35055514} (\bibinfo {year} {2001})\BibitemShut
  {NoStop}%
\bibitem [{\citenamefont {Sciortino}(2011)}]{Sciorino_Silicon_2011}%
  \BibitemOpen
  \bibfield  {author} {\bibinfo {author} {\bibfnamefont {F.}~\bibnamefont
  {Sciortino}},\ }\bibfield  {title} {\bibinfo {title} {Liquid–liquid
  transitions: Silicon in silico},\ }\href
  {https://doi.org/https://doi.org/10.1038/nphys2038} {\bibfield  {journal}
  {\bibinfo  {journal} {Nat. Phys.}\ }\textbf {\bibinfo {volume} {7}},\
  \bibinfo {pages} {523} (\bibinfo {year} {2011})}\BibitemShut {NoStop}%
\bibitem [{\citenamefont {Ohta}\ \emph {et~al.}(2015)\citenamefont {Ohta},
  \citenamefont {Ichimaru}, \citenamefont {Einaga}, \citenamefont {Kawaguchi},
  \citenamefont {Shimizu}, \citenamefont {Matsuoka}, \citenamefont {Hirao},\
  and\ \citenamefont {Ohishi}}]{Ohta_H_2015}%
  \BibitemOpen
  \bibfield  {author} {\bibinfo {author} {\bibfnamefont {K.}~\bibnamefont
  {Ohta}}, \bibinfo {author} {\bibfnamefont {K.}~\bibnamefont {Ichimaru}},
  \bibinfo {author} {\bibfnamefont {M.}~\bibnamefont {Einaga}}, \bibinfo
  {author} {\bibfnamefont {S.}~\bibnamefont {Kawaguchi}}, \bibinfo {author}
  {\bibfnamefont {K.}~\bibnamefont {Shimizu}}, \bibinfo {author} {\bibfnamefont
  {T.}~\bibnamefont {Matsuoka}}, \bibinfo {author} {\bibfnamefont
  {N.}~\bibnamefont {Hirao}},\ and\ \bibinfo {author} {\bibfnamefont
  {Y.}~\bibnamefont {Ohishi}},\ }\bibfield  {title} {\bibinfo {title} {Phase
  boundary of hot dense fluid hydrogen},\ }\bibfield  {journal} {\bibinfo
  {journal} {Sci. Rep.}\ }\textbf {\bibinfo {volume} {5}},\ \href
  {https://doi.org/https://doi.org/10.1038/srep16560}
  {https://doi.org/10.1038/srep16560} (\bibinfo {year} {2015})\BibitemShut
  {NoStop}%
\bibitem [{\citenamefont {McWilliams}\ \emph {et~al.}(2016)\citenamefont
  {McWilliams}, \citenamefont {Dalton}, \citenamefont {Mahmood},\ and\
  \citenamefont {Goncharov}}]{McWilliams_H_2016}%
  \BibitemOpen
  \bibfield  {author} {\bibinfo {author} {\bibfnamefont {R.~S.}\ \bibnamefont
  {McWilliams}}, \bibinfo {author} {\bibfnamefont {D.~A.}\ \bibnamefont
  {Dalton}}, \bibinfo {author} {\bibfnamefont {M.~F.}\ \bibnamefont
  {Mahmood}},\ and\ \bibinfo {author} {\bibfnamefont {A.~F.}\ \bibnamefont
  {Goncharov}},\ }\bibfield  {title} {\bibinfo {title} {Optical properties of
  fluid hydrogen at the transition to a conducting state},\ }\href
  {https://doi.org/10.1103/PhysRevLett.116.255501} {\bibfield  {journal}
  {\bibinfo  {journal} {Phys. Rev. Lett.}\ }\textbf {\bibinfo {volume} {116}},\
  \bibinfo {pages} {255501} (\bibinfo {year} {2016})}\BibitemShut {NoStop}%
\bibitem [{\citenamefont {Dzyabura}\ \emph {et~al.}(2013)\citenamefont
  {Dzyabura}, \citenamefont {Zaghoo},\ and\ \citenamefont
  {Silvera}}]{Zaghoo_H_2013}%
  \BibitemOpen
  \bibfield  {author} {\bibinfo {author} {\bibfnamefont {V.}~\bibnamefont
  {Dzyabura}}, \bibinfo {author} {\bibfnamefont {M.}~\bibnamefont {Zaghoo}},\
  and\ \bibinfo {author} {\bibfnamefont {I.~F.}\ \bibnamefont {Silvera}},\
  }\bibfield  {title} {\bibinfo {title} {Evidence of a liquid\&\#x2013;liquid
  phase transition in hot dense hydrogen},\ }\href
  {https://doi.org/10.1073/pnas.1300718110} {\bibfield  {journal} {\bibinfo
  {journal} {Proc. Natl. Acad. Sci.}\ }\textbf {\bibinfo {volume} {110}},\
  \bibinfo {pages} {8040} (\bibinfo {year} {2013})}\BibitemShut {NoStop}%
\bibitem [{\citenamefont {Zaghoo}\ \emph {et~al.}(2016)\citenamefont {Zaghoo},
  \citenamefont {Salamat},\ and\ \citenamefont {Silvera}}]{Zaghoo_H_2016}%
  \BibitemOpen
  \bibfield  {author} {\bibinfo {author} {\bibfnamefont {M.}~\bibnamefont
  {Zaghoo}}, \bibinfo {author} {\bibfnamefont {A.}~\bibnamefont {Salamat}},\
  and\ \bibinfo {author} {\bibfnamefont {I.~F.}\ \bibnamefont {Silvera}},\
  }\bibfield  {title} {\bibinfo {title} {Evidence of a first-order phase
  transition to metallic hydrogen},\ }\href
  {https://doi.org/10.1103/PhysRevB.93.155128} {\bibfield  {journal} {\bibinfo
  {journal} {Phys. Rev. B}\ }\textbf {\bibinfo {volume} {93}},\ \bibinfo
  {pages} {155128} (\bibinfo {year} {2016})}\BibitemShut {NoStop}%
\bibitem [{\citenamefont {Zaghoo}\ and\ \citenamefont
  {Silvera}(2017)}]{Zaghoo_H_2017}%
  \BibitemOpen
  \bibfield  {author} {\bibinfo {author} {\bibfnamefont {M.}~\bibnamefont
  {Zaghoo}}\ and\ \bibinfo {author} {\bibfnamefont {I.~F.}\ \bibnamefont
  {Silvera}},\ }\bibfield  {title} {\bibinfo {title} {Conductivity and
  dissociation in liquid metallic hydrogen and implications for planetary
  interiors},\ }\href {https://doi.org/10.1073/pnas.1707918114} {\bibfield
  {journal} {\bibinfo  {journal} {Proc. Natl. Acad. Sci.}\ }\textbf {\bibinfo
  {volume} {114}},\ \bibinfo {pages} {11873} (\bibinfo {year}
  {2017})}\BibitemShut {NoStop}%
\bibitem [{\citenamefont {Vollhardt}\ and\ \citenamefont
  {Wölfle}(1990)}]{Vollhardt_He_1990}%
  \BibitemOpen
  \bibfield  {author} {\bibinfo {author} {\bibfnamefont {D.}~\bibnamefont
  {Vollhardt}}\ and\ \bibinfo {author} {\bibfnamefont {P.}~\bibnamefont
  {Wölfle}},\ }\href@noop {} {\emph {\bibinfo {title} {The Superfluid Phases
  of Helium 3}}}\ (\bibinfo  {publisher} {Taylor and Francis},\ \bibinfo
  {address} {London, UK},\ \bibinfo {year} {1990})\BibitemShut {NoStop}%
\bibitem [{\citenamefont {Schmitt}(2015)}]{Schmitt_He_2015}%
  \BibitemOpen
  \bibfield  {author} {\bibinfo {author} {\bibfnamefont {A.}~\bibnamefont
  {Schmitt}},\ }\bibfield  {title} {\bibinfo {title} {Introduction to
  superfluidity}\ }(\bibinfo  {publisher} {Springer International Publishing},\
  \bibinfo {address} {Cham},\ \bibinfo {year} {2015})\BibitemShut {NoStop}%
\bibitem [{\citenamefont {Henry}\ \emph {et~al.}(2020)\citenamefont {Henry},
  \citenamefont {Mezouar}, \citenamefont {Garbarino}, \citenamefont {Sifré},
  \citenamefont {Weck},\ and\ \citenamefont {Datchi}}]{Henry2020}%
  \BibitemOpen
  \bibfield  {author} {\bibinfo {author} {\bibfnamefont {L.}~\bibnamefont
  {Henry}}, \bibinfo {author} {\bibfnamefont {M.}~\bibnamefont {Mezouar}},
  \bibinfo {author} {\bibfnamefont {G.}~\bibnamefont {Garbarino}}, \bibinfo
  {author} {\bibfnamefont {D.}~\bibnamefont {Sifré}}, \bibinfo {author}
  {\bibfnamefont {G.}~\bibnamefont {Weck}},\ and\ \bibinfo {author}
  {\bibfnamefont {F.}~\bibnamefont {Datchi}},\ }\bibfield  {title} {\bibinfo
  {title} {Liquid–liquid transition and critical point in sulfur},\ }\href
  {https://doi.org/https://doi.org/10.1038/s41586-020-2593-1} {\bibfield
  {journal} {\bibinfo  {journal} {Nature}\ }\textbf {\bibinfo {volume} {584}},\
  \bibinfo {pages} {382} (\bibinfo {year} {2020})}\BibitemShut {NoStop}%
\bibitem [{\citenamefont {Katayama}\ \emph {et~al.}(2000)\citenamefont
  {Katayama}, \citenamefont {Mizutani}, \citenamefont {Utsumi}, \citenamefont
  {Shimomura}, \citenamefont {Yamakata},\ and\ \citenamefont {ichi
  Funakoshi}}]{Katayama2000}%
  \BibitemOpen
  \bibfield  {author} {\bibinfo {author} {\bibfnamefont {Y.}~\bibnamefont
  {Katayama}}, \bibinfo {author} {\bibfnamefont {T.}~\bibnamefont {Mizutani}},
  \bibinfo {author} {\bibfnamefont {W.}~\bibnamefont {Utsumi}}, \bibinfo
  {author} {\bibfnamefont {O.}~\bibnamefont {Shimomura}}, \bibinfo {author}
  {\bibfnamefont {M.}~\bibnamefont {Yamakata}},\ and\ \bibinfo {author}
  {\bibfnamefont {K.}~\bibnamefont {ichi Funakoshi}},\ }\bibfield  {title}
  {\bibinfo {title} {A first-order liquid–liquid phase transition in
  phosphorus},\ }\href {https://doi.org/https://doi.org/10.1038/35003143}
  {\bibfield  {journal} {\bibinfo  {journal} {Nature}\ }\textbf {\bibinfo
  {volume} {403}},\ \bibinfo {pages} {170} (\bibinfo {year}
  {2000})}\BibitemShut {NoStop}%
\bibitem [{\citenamefont {Katayama}\ \emph {et~al.}(2004)\citenamefont
  {Katayama}, \citenamefont {Inamura}, \citenamefont {Mizutani}, \citenamefont
  {Yamakata}, \citenamefont {Utsumi},\ and\ \citenamefont
  {Shimomura}}]{Katayama_Phos2_2004}%
  \BibitemOpen
  \bibfield  {author} {\bibinfo {author} {\bibfnamefont {Y.}~\bibnamefont
  {Katayama}}, \bibinfo {author} {\bibfnamefont {Y.}~\bibnamefont {Inamura}},
  \bibinfo {author} {\bibfnamefont {T.}~\bibnamefont {Mizutani}}, \bibinfo
  {author} {\bibfnamefont {M.}~\bibnamefont {Yamakata}}, \bibinfo {author}
  {\bibfnamefont {W.}~\bibnamefont {Utsumi}},\ and\ \bibinfo {author}
  {\bibfnamefont {O.}~\bibnamefont {Shimomura}},\ }\bibfield  {title} {\bibinfo
  {title} {Macroscopic separation of dense fluid phase and liquid phase of
  phosphorus},\ }\href {https://doi.org/10.1126/science.1102735} {\bibfield
  {journal} {\bibinfo  {journal} {Science}\ }\textbf {\bibinfo {volume}
  {306}},\ \bibinfo {pages} {848} (\bibinfo {year} {2004})}\BibitemShut
  {NoStop}%
\bibitem [{\citenamefont {Glosli}\ and\ \citenamefont
  {Ree}(1999)}]{Glosli_Liquid_1999}%
  \BibitemOpen
  \bibfield  {author} {\bibinfo {author} {\bibfnamefont {J.~N.}\ \bibnamefont
  {Glosli}}\ and\ \bibinfo {author} {\bibfnamefont {F.~H.}\ \bibnamefont
  {Ree}},\ }\bibfield  {title} {\bibinfo {title} {Liquid-liquid phase
  transformation in carbon},\ }\href
  {https://doi.org/10.1103/PhysRevLett.82.4659} {\bibfield  {journal} {\bibinfo
   {journal} {Phys. Rev. Lett.}\ }\textbf {\bibinfo {volume} {82}},\ \bibinfo
  {pages} {4659} (\bibinfo {year} {1999})}\BibitemShut {NoStop}%
\bibitem [{\citenamefont {Brazhkin}\ \emph {et~al.}(1999)\citenamefont
  {Brazhkin}, \citenamefont {Popova},\ and\ \citenamefont
  {Voloshin}}]{Brazhkin_PT_1999}%
  \BibitemOpen
  \bibfield  {author} {\bibinfo {author} {\bibfnamefont {V.~V.}\ \bibnamefont
  {Brazhkin}}, \bibinfo {author} {\bibfnamefont {S.~V.}\ \bibnamefont
  {Popova}},\ and\ \bibinfo {author} {\bibfnamefont {R.~N.}\ \bibnamefont
  {Voloshin}},\ }\bibfield  {title} {\bibinfo {title} {Pressure -temperature
  phase diagram of molten elements: selenium, sulfur and iodine},\ }\href
  {https://doi.org/https://doi.org/10.1016/S0921-4526(98)01318-0} {\bibfield
  {journal} {\bibinfo  {journal} {Physica B}\ }\textbf {\bibinfo {volume}
  {265}},\ \bibinfo {pages} {64} (\bibinfo {year} {1999})}\BibitemShut
  {NoStop}%
\bibitem [{\citenamefont {Plašienka}\ \emph {et~al.}(2015)\citenamefont
  {Plašienka}, \citenamefont {Cifra},\ and\ \citenamefont
  {Martoňák}}]{Plasienka_Structural_2015}%
  \BibitemOpen
  \bibfield  {author} {\bibinfo {author} {\bibfnamefont {D.}~\bibnamefont
  {Plašienka}}, \bibinfo {author} {\bibfnamefont {P.}~\bibnamefont {Cifra}},\
  and\ \bibinfo {author} {\bibfnamefont {R.}~\bibnamefont {Martoňák}},\
  }\bibfield  {title} {\bibinfo {title} {Structural transformation between long
  and short-chain form of liquid sulfur from ab initio molecular dynamics},\
  }\href {https://doi.org/https://doi.org/10.1063/1.4917040} {\bibfield
  {journal} {\bibinfo  {journal} {J. Chem. Phys.}\ }\textbf {\bibinfo {volume}
  {142}},\ \bibinfo {pages} {154502} (\bibinfo {year} {2015})}\BibitemShut
  {NoStop}%
\bibitem [{\citenamefont {Saika-Voivod}\ \emph {et~al.}(2004)\citenamefont
  {Saika-Voivod}, \citenamefont {Sciortino}, \citenamefont {Grande},\ and\
  \citenamefont {Poole}}]{Saika_Silica_2004}%
  \BibitemOpen
  \bibfield  {author} {\bibinfo {author} {\bibfnamefont {I.}~\bibnamefont
  {Saika-Voivod}}, \bibinfo {author} {\bibfnamefont {F.}~\bibnamefont
  {Sciortino}}, \bibinfo {author} {\bibfnamefont {T.}~\bibnamefont {Grande}},\
  and\ \bibinfo {author} {\bibfnamefont {P.~H.}\ \bibnamefont {Poole}},\
  }\bibfield  {title} {\bibinfo {title} {Phase diagram of silica from computer
  simulation},\ }\href {https://doi.org/10.1103/PhysRevE.70.061507} {\bibfield
  {journal} {\bibinfo  {journal} {Phys. Rev. E}\ }\textbf {\bibinfo {volume}
  {70}},\ \bibinfo {pages} {061507} (\bibinfo {year} {2004})}\BibitemShut
  {NoStop}%
\bibitem [{\citenamefont {Lascaris}\ \emph {et~al.}(2014)\citenamefont
  {Lascaris}, \citenamefont {Hemmati}, \citenamefont {Buldyrev}, \citenamefont
  {Stanley},\ and\ \citenamefont {Angell}}]{Lascaris_Silica_2014}%
  \BibitemOpen
  \bibfield  {author} {\bibinfo {author} {\bibfnamefont {E.}~\bibnamefont
  {Lascaris}}, \bibinfo {author} {\bibfnamefont {M.}~\bibnamefont {Hemmati}},
  \bibinfo {author} {\bibfnamefont {S.~V.}\ \bibnamefont {Buldyrev}}, \bibinfo
  {author} {\bibfnamefont {H.~E.}\ \bibnamefont {Stanley}},\ and\ \bibinfo
  {author} {\bibfnamefont {C.~A.}\ \bibnamefont {Angell}},\ }\bibfield  {title}
  {\bibinfo {title} {Search for a liquid-liquid critical point in models of
  silica},\ }\href {https://doi.org/10.1063/1.4879057} {\bibfield  {journal}
  {\bibinfo  {journal} {J. Chem. Phys.}\ }\textbf {\bibinfo {volume} {140}},\
  \bibinfo {pages} {224502} (\bibinfo {year} {2014})}\BibitemShut {NoStop}%
\bibitem [{\citenamefont {Chen}\ \emph {et~al.}(2017)\citenamefont {Chen},
  \citenamefont {Lascaris}, ,\ and\ \citenamefont {Palmer}}]{Chen_Silica_2017}%
  \BibitemOpen
  \bibfield  {author} {\bibinfo {author} {\bibfnamefont {R.}~\bibnamefont
  {Chen}}, \bibinfo {author} {\bibfnamefont {E.}~\bibnamefont {Lascaris}}, ,\
  and\ \bibinfo {author} {\bibfnamefont {J.~C.}\ \bibnamefont {Palmer}},\
  }\bibfield  {title} {\bibinfo {title} {Liquid–liquid phase transition in an
  ionic model of silica},\ }\href
  {https://doi.org/https://doi.org/10.1063/1.4984335} {\bibfield  {journal}
  {\bibinfo  {journal} {J. Chem. Phys.}\ }\textbf {\bibinfo {volume} {146}},\
  \bibinfo {pages} {234503} (\bibinfo {year} {2017})}\BibitemShut {NoStop}%
\bibitem [{\citenamefont {Sastry}\ and\ \citenamefont
  {Angell}(2003)}]{Sastry_Silicon_2003}%
  \BibitemOpen
  \bibfield  {author} {\bibinfo {author} {\bibfnamefont {S.}~\bibnamefont
  {Sastry}}\ and\ \bibinfo {author} {\bibfnamefont {C.~A.}\ \bibnamefont
  {Angell}},\ }\bibfield  {title} {\bibinfo {title} {Liquid–liquid phase
  transition in supercooled silicon},\ }\href {https://doi.org/10.1038/nmat994}
  {\bibfield  {journal} {\bibinfo  {journal} {Nat. Mater.}\ }\textbf {\bibinfo
  {volume} {2}},\ \bibinfo {pages} {739} (\bibinfo {year} {2003})}\BibitemShut
  {NoStop}%
\bibitem [{\citenamefont {Holten}\ and\ \citenamefont
  {Anisimov}(2012)}]{Holten_Liquid_2012}%
  \BibitemOpen
  \bibfield  {author} {\bibinfo {author} {\bibfnamefont {V.}~\bibnamefont
  {Holten}}\ and\ \bibinfo {author} {\bibfnamefont {M.~A.}\ \bibnamefont
  {Anisimov}},\ }\bibfield  {title} {\bibinfo {title} {Entropy-driven
  liquid–liquid separation in supercooled water},\ }\href
  {https://doi.org/https://doi.org/10.1038/srep00713} {\bibfield  {journal}
  {\bibinfo  {journal} {Sci. Rep.}\ }\textbf {\bibinfo {volume} {2}},\ \bibinfo
  {pages} {713} (\bibinfo {year} {2012})}\BibitemShut {NoStop}%
\bibitem [{\citenamefont {Gallo}\ \emph {et~al.}(2016)\citenamefont {Gallo},
  \citenamefont {Amann-Winkel}, \citenamefont {Angell}, \citenamefont
  {Anisimov}, \citenamefont {Caupin}, \citenamefont {Chakravarty},
  \citenamefont {Lascaris}, \citenamefont {Loerting}, \citenamefont
  {Panagiotopoulos}, \citenamefont {Russo}, \citenamefont {Sellberg},
  \citenamefont {Stanley}, \citenamefont {Tanaka}, \citenamefont {Vega},
  \citenamefont {Xu},\ and\ \citenamefont {Pettersson}}]{Gallo2016}%
  \BibitemOpen
  \bibfield  {author} {\bibinfo {author} {\bibfnamefont {P.}~\bibnamefont
  {Gallo}}, \bibinfo {author} {\bibfnamefont {K.}~\bibnamefont {Amann-Winkel}},
  \bibinfo {author} {\bibfnamefont {C.~A.}\ \bibnamefont {Angell}}, \bibinfo
  {author} {\bibfnamefont {M.~A.}\ \bibnamefont {Anisimov}}, \bibinfo {author}
  {\bibfnamefont {F.}~\bibnamefont {Caupin}}, \bibinfo {author} {\bibfnamefont
  {C.}~\bibnamefont {Chakravarty}}, \bibinfo {author} {\bibfnamefont
  {E.}~\bibnamefont {Lascaris}}, \bibinfo {author} {\bibfnamefont
  {T.}~\bibnamefont {Loerting}}, \bibinfo {author} {\bibfnamefont {A.~Z.}\
  \bibnamefont {Panagiotopoulos}}, \bibinfo {author} {\bibfnamefont
  {J.}~\bibnamefont {Russo}}, \bibinfo {author} {\bibfnamefont {J.~A.}\
  \bibnamefont {Sellberg}}, \bibinfo {author} {\bibfnamefont {H.~E.}\
  \bibnamefont {Stanley}}, \bibinfo {author} {\bibfnamefont {H.}~\bibnamefont
  {Tanaka}}, \bibinfo {author} {\bibfnamefont {C.}~\bibnamefont {Vega}},
  \bibinfo {author} {\bibfnamefont {L.}~\bibnamefont {Xu}},\ and\ \bibinfo
  {author} {\bibfnamefont {L.~G.~M.}\ \bibnamefont {Pettersson}},\ }\bibfield
  {title} {\bibinfo {title} {Water: A tale of two liquids},\ }\href
  {https://doi.org/https://doi.org/10.1021/acs.chemrev.5b00750} {\bibfield
  {journal} {\bibinfo  {journal} {Chem. Rev.}\ }\textbf {\bibinfo {volume}
  {116}},\ \bibinfo {pages} {7463} (\bibinfo {year} {2016})}\BibitemShut
  {NoStop}%
\bibitem [{\citenamefont {Duška}(2020)}]{Duska_Water_2020}%
  \BibitemOpen
  \bibfield  {author} {\bibinfo {author} {\bibfnamefont {M.}~\bibnamefont
  {Duška}},\ }\bibfield  {title} {\bibinfo {title} {Water above the
  spinodal},\ }\href {https://doi.org/https://doi.org/10.1063/5.0006431}
  {\bibfield  {journal} {\bibinfo  {journal} {J. Chem. Phys.}\ }\textbf
  {\bibinfo {volume} {152}},\ \bibinfo {pages} {174501} (\bibinfo {year}
  {2020})}\BibitemShut {NoStop}%
\bibitem [{\citenamefont {Caupin}\ and\ \citenamefont
  {Anisimov}(2019)}]{Caupin_Thermodynamics_2019}%
  \BibitemOpen
  \bibfield  {author} {\bibinfo {author} {\bibfnamefont {F.}~\bibnamefont
  {Caupin}}\ and\ \bibinfo {author} {\bibfnamefont {M.~A.}\ \bibnamefont
  {Anisimov}},\ }\bibfield  {title} {\bibinfo {title} {Thermodynamics of
  supercooled and stretched water: Unifying two-structure description and
  liquid-vapor spinodal},\ }\href
  {https://doi.org/https://doi.org/10.1063/1.5100228} {\bibfield  {journal}
  {\bibinfo  {journal} {J. Chem. Phys.}\ }\textbf {\bibinfo {volume} {151}},\
  \bibinfo {pages} {034503} (\bibinfo {year} {2019})}\BibitemShut {NoStop}%
\bibitem [{\citenamefont {Poole}\ \emph {et~al.}(1992)\citenamefont {Poole},
  \citenamefont {Sciortino}, \citenamefont {Essmann},\ and\ \citenamefont
  {Stanley}}]{Poole1992}%
  \BibitemOpen
  \bibfield  {author} {\bibinfo {author} {\bibfnamefont {P.~H.}\ \bibnamefont
  {Poole}}, \bibinfo {author} {\bibfnamefont {F.}~\bibnamefont {Sciortino}},
  \bibinfo {author} {\bibfnamefont {U.}~\bibnamefont {Essmann}},\ and\ \bibinfo
  {author} {\bibfnamefont {H.~E.}\ \bibnamefont {Stanley}},\ }\bibfield
  {title} {\bibinfo {title} {Phase behavior of metastable water},\ }\href
  {https://doi.org/https://doi.org/10.1038/360324a0} {\bibfield  {journal}
  {\bibinfo  {journal} {Nature}\ }\textbf {\bibinfo {volume} {360}},\ \bibinfo
  {pages} {324} (\bibinfo {year} {1992})}\BibitemShut {NoStop}%
\bibitem [{\citenamefont {Holten}\ \emph {et~al.}(2014)\citenamefont {Holten},
  \citenamefont {Palmer}, \citenamefont {Poole}, \citenamefont {Debenedetti},\
  and\ \citenamefont {Anisimov}}]{Holten2001}%
  \BibitemOpen
  \bibfield  {author} {\bibinfo {author} {\bibfnamefont {V.}~\bibnamefont
  {Holten}}, \bibinfo {author} {\bibfnamefont {J.~C.}\ \bibnamefont {Palmer}},
  \bibinfo {author} {\bibfnamefont {P.~H.}\ \bibnamefont {Poole}}, \bibinfo
  {author} {\bibfnamefont {P.~G.}\ \bibnamefont {Debenedetti}},\ and\ \bibinfo
  {author} {\bibfnamefont {M.~A.}\ \bibnamefont {Anisimov}},\ }\bibfield
  {title} {\bibinfo {title} {Two-state thermodynamics of the st2 model for
  supercooled water},\ }\bibfield  {journal} {\bibinfo  {journal} {J. Chem.
  Phys.}\ }\textbf {\bibinfo {volume} {104502}},\ \href
  {https://doi.org/https://doi.org/10.1063/1.4867287}
  {https://doi.org/10.1063/1.4867287} (\bibinfo {year} {2014})\BibitemShut
  {NoStop}%
\bibitem [{\citenamefont {Debenedetti}\ \emph {et~al.}(2020)\citenamefont
  {Debenedetti}, \citenamefont {Sciortino},\ and\ \citenamefont
  {Zerze}}]{Debenedetti2020}%
  \BibitemOpen
  \bibfield  {author} {\bibinfo {author} {\bibfnamefont {P.~G.}\ \bibnamefont
  {Debenedetti}}, \bibinfo {author} {\bibfnamefont {F.}~\bibnamefont
  {Sciortino}},\ and\ \bibinfo {author} {\bibfnamefont {G.~H.}\ \bibnamefont
  {Zerze}},\ }\bibfield  {title} {\bibinfo {title} {Second critical point in
  two realistic models of water},\ }\href
  {https://doi.org/10.1126/science.abb9796} {\bibfield  {journal} {\bibinfo
  {journal} {Science}\ }\textbf {\bibinfo {volume} {369}},\ \bibinfo {pages}
  {289} (\bibinfo {year} {2020})}\BibitemShut {NoStop}%
\bibitem [{\citenamefont {Biddle}\ \emph {et~al.}(2017)\citenamefont {Biddle},
  \citenamefont {Singh}, \citenamefont {Sparano}, \citenamefont {Ricci},
  \citenamefont {González}, \citenamefont {Valeriani}, \citenamefont
  {Abascal}, \citenamefont {Debenedetti}, \citenamefont {Anisimov}, ,\ and\
  \citenamefont {Caupin}}]{Biddle_Two_2017}%
  \BibitemOpen
  \bibfield  {author} {\bibinfo {author} {\bibfnamefont {J.~W.}\ \bibnamefont
  {Biddle}}, \bibinfo {author} {\bibfnamefont {R.~S.}\ \bibnamefont {Singh}},
  \bibinfo {author} {\bibfnamefont {E.~M.}\ \bibnamefont {Sparano}}, \bibinfo
  {author} {\bibfnamefont {F.}~\bibnamefont {Ricci}}, \bibinfo {author}
  {\bibfnamefont {M.~A.}\ \bibnamefont {González}}, \bibinfo {author}
  {\bibfnamefont {C.}~\bibnamefont {Valeriani}}, \bibinfo {author}
  {\bibfnamefont {J.~L.~F.}\ \bibnamefont {Abascal}}, \bibinfo {author}
  {\bibfnamefont {P.~G.}\ \bibnamefont {Debenedetti}}, \bibinfo {author}
  {\bibfnamefont {M.~A.}\ \bibnamefont {Anisimov}}, ,\ and\ \bibinfo {author}
  {\bibfnamefont {F.}~\bibnamefont {Caupin}},\ }\bibfield  {title} {\bibinfo
  {title} {Two-structure thermodynamics for the tip4p/2005 model of water
  covering supercooled and deeply stretched regions},\ }\href
  {https://doi.org/10.1063/1.4973546} {\bibfield  {journal} {\bibinfo
  {journal} {J. Chem. Phys.}\ }\textbf {\bibinfo {volume} {146}},\ \bibinfo
  {pages} {034502} (\bibinfo {year} {2017})}\BibitemShut {NoStop}%
\bibitem [{\citenamefont {Debenedetti}(1998)}]{Debenedetti_One_1998}%
  \BibitemOpen
  \bibfield  {author} {\bibinfo {author} {\bibfnamefont {P.~G.}\ \bibnamefont
  {Debenedetti}},\ }\bibfield  {title} {\bibinfo {title} {One substance, two
  liquids?},\ }\href {https://doi.org/https://doi.org/10.1038/32286} {\bibfield
   {journal} {\bibinfo  {journal} {Nature}\ }\textbf {\bibinfo {volume}
  {392}},\ \bibinfo {pages} {127} (\bibinfo {year} {1998})}\BibitemShut
  {NoStop}%
\bibitem [{\citenamefont {Longo}\ and\ \citenamefont
  {Anisimov}(2022)}]{Longo2021}%
  \BibitemOpen
  \bibfield  {author} {\bibinfo {author} {\bibfnamefont {T.~J.}\ \bibnamefont
  {Longo}}\ and\ \bibinfo {author} {\bibfnamefont {M.~A.}\ \bibnamefont
  {Anisimov}},\ }\bibfield  {title} {\bibinfo {title} {Phase transitions
  affected by natural and forceful molecular interconversion},\ }\href
  {https://doi.org/https://doi.org/10.1063/5.0081180} {\bibfield  {journal}
  {\bibinfo  {journal} {J. Chem. Phys.}\ }\textbf {\bibinfo {volume} {156}},\
  \bibinfo {pages} {084502} (\bibinfo {year} {2022})}\BibitemShut {NoStop}%
\bibitem [{\citenamefont {Caupin}\ and\ \citenamefont
  {Anisimov}(2021)}]{Caupin2021}%
  \BibitemOpen
  \bibfield  {author} {\bibinfo {author} {\bibfnamefont {F.}~\bibnamefont
  {Caupin}}\ and\ \bibinfo {author} {\bibfnamefont {M.~A.}\ \bibnamefont
  {Anisimov}},\ }\bibfield  {title} {\bibinfo {title} {Minimal microscopic
  model for liquid polyamorphism and waterlike anomalies},\ }\href
  {https://doi.org/10.1103/PhysRevLett.127.185701} {\bibfield  {journal}
  {\bibinfo  {journal} {Phys. Rev. Lett.}\ }\textbf {\bibinfo {volume} {127}},\
  \bibinfo {pages} {185701} (\bibinfo {year} {2021})}\BibitemShut {NoStop}%
\bibitem [{\citenamefont {Sauer}\ and\ \citenamefont
  {Borst}(1967)}]{Sauer_Lambda_1967}%
  \BibitemOpen
  \bibfield  {author} {\bibinfo {author} {\bibfnamefont {G.~E.}\ \bibnamefont
  {Sauer}}\ and\ \bibinfo {author} {\bibfnamefont {L.~B.}\ \bibnamefont
  {Borst}},\ }\bibfield  {title} {\bibinfo {title} {Lambda transition in liquid
  sulfur},\ }\href {https://doi.org/10.1126/science.158.3808.1567} {\bibfield
  {journal} {\bibinfo  {journal} {Science}\ }\textbf {\bibinfo {volume}
  {158}},\ \bibinfo {pages} {1567} (\bibinfo {year} {1967})}\BibitemShut
  {NoStop}%
\bibitem [{\citenamefont {Bellissent}\ \emph {et~al.}(1994)\citenamefont
  {Bellissent}, \citenamefont {Descotes},\ and\ \citenamefont
  {Pfeuty}}]{Bellissent_Sulfur_1994}%
  \BibitemOpen
  \bibfield  {author} {\bibinfo {author} {\bibfnamefont {R.}~\bibnamefont
  {Bellissent}}, \bibinfo {author} {\bibfnamefont {L.}~\bibnamefont
  {Descotes}},\ and\ \bibinfo {author} {\bibfnamefont {P.}~\bibnamefont
  {Pfeuty}},\ }\bibfield  {title} {\bibinfo {title} {Polymerization in liquid
  sulphur},\ }\href {https://doi.org/10.1088/0953-8984/6/23a/031} {\bibfield
  {journal} {\bibinfo  {journal} {J. Phys.: Condens. Matte}\ }\textbf {\bibinfo
  {volume} {6}},\ \bibinfo {pages} {A211} (\bibinfo {year} {1994})}\BibitemShut
  {NoStop}%
\bibitem [{\citenamefont {Kozhevnikov}\ \emph {et~al.}(2004)\citenamefont
  {Kozhevnikov}, \citenamefont {Payne}, \citenamefont {Olson}, \citenamefont
  {McDonald},\ and\ \citenamefont {Inglefield}}]{Kozhevnikov_Sulfur_2004}%
  \BibitemOpen
  \bibfield  {author} {\bibinfo {author} {\bibfnamefont {V.~F.}\ \bibnamefont
  {Kozhevnikov}}, \bibinfo {author} {\bibfnamefont {W.~B.}\ \bibnamefont
  {Payne}}, \bibinfo {author} {\bibfnamefont {J.~K.}\ \bibnamefont {Olson}},
  \bibinfo {author} {\bibfnamefont {C.~L.}\ \bibnamefont {McDonald}},\ and\
  \bibinfo {author} {\bibfnamefont {C.~E.}\ \bibnamefont {Inglefield}},\
  }\bibfield  {title} {\bibinfo {title} {Physical properties of sulfur near the
  polymerization transition},\ }\href {https://doi.org/10.1063/1.1794031}
  {\bibfield  {journal} {\bibinfo  {journal} {J. Chem. Phys.}\ }\textbf
  {\bibinfo {volume} {121}} (\bibinfo {year} {2004})}\BibitemShut {NoStop}%
\bibitem [{\citenamefont {Tobolsky}\ and\ \citenamefont
  {Eisenberg}(1959)}]{Tobolsky_Sulfur_1959}%
  \BibitemOpen
  \bibfield  {author} {\bibinfo {author} {\bibfnamefont {A.~V.}\ \bibnamefont
  {Tobolsky}}\ and\ \bibinfo {author} {\bibfnamefont {A.}~\bibnamefont
  {Eisenberg}},\ }\bibfield  {title} {\bibinfo {title} {Equilibrium
  polymerization of sulfur},\ }\href
  {https://doi.org/https://doi.org/10.1021/ja01513a004} {\bibfield  {journal}
  {\bibinfo  {journal} {J. Am. Chem. Soc.}\ }\textbf {\bibinfo {volume} {81}},\
  \bibinfo {pages} {780} (\bibinfo {year} {1959})}\BibitemShut {NoStop}%
\bibitem [{\citenamefont {Eisenberg}\ and\ \citenamefont
  {Tobolsky}(1960)}]{Tobolsky_Selenium_1960}%
  \BibitemOpen
  \bibfield  {author} {\bibinfo {author} {\bibfnamefont {A.}~\bibnamefont
  {Eisenberg}}\ and\ \bibinfo {author} {\bibfnamefont {A.~V.}\ \bibnamefont
  {Tobolsky}},\ }\bibfield  {title} {\bibinfo {title} {Equilibrium
  polymerization of selenium},\ }\bibfield  {journal} {\bibinfo  {journal} {J.
  Pol. Sci.}\ }\textbf {\bibinfo {volume} {46}},\ \href
  {https://doi.org/https://doi.org/10.1002/pol.1960.1204614703}
  {https://doi.org/10.1002/pol.1960.1204614703} (\bibinfo {year}
  {1960})\BibitemShut {NoStop}%
\bibitem [{\citenamefont {Morales}\ \emph {et~al.}(2010)\citenamefont
  {Morales}, \citenamefont {Pierleoni}, \citenamefont {Schwegler},\ and\
  \citenamefont {Ceperley}}]{Morales_H_2010}%
  \BibitemOpen
  \bibfield  {author} {\bibinfo {author} {\bibfnamefont {M.~A.}\ \bibnamefont
  {Morales}}, \bibinfo {author} {\bibfnamefont {C.}~\bibnamefont {Pierleoni}},
  \bibinfo {author} {\bibfnamefont {E.}~\bibnamefont {Schwegler}},\ and\
  \bibinfo {author} {\bibfnamefont {D.~M.}\ \bibnamefont {Ceperley}},\
  }\bibfield  {title} {\bibinfo {title} {Evidence for a first-order
  liquid-liquid transition in high-pressure hydrogen from ab initio
  simulations},\ }\href {https://doi.org/10.1073/pnas.1007309107} {\bibfield
  {journal} {\bibinfo  {journal} {Proc. Natl. Acad. Sci.}\ }\textbf {\bibinfo
  {volume} {107}},\ \bibinfo {pages} {12799} (\bibinfo {year}
  {2010})}\BibitemShut {NoStop}%
\bibitem [{\citenamefont {Pierleoni}\ \emph {et~al.}(2016)\citenamefont
  {Pierleoni}, \citenamefont {Morales}, \citenamefont {Rillo}, \citenamefont
  {Holzmann},\ and\ \citenamefont {Ceperley}}]{Pierleoni_H_2016}%
  \BibitemOpen
  \bibfield  {author} {\bibinfo {author} {\bibfnamefont {C.}~\bibnamefont
  {Pierleoni}}, \bibinfo {author} {\bibfnamefont {M.~A.}\ \bibnamefont
  {Morales}}, \bibinfo {author} {\bibfnamefont {G.}~\bibnamefont {Rillo}},
  \bibinfo {author} {\bibfnamefont {M.}~\bibnamefont {Holzmann}},\ and\
  \bibinfo {author} {\bibfnamefont {D.~M.}\ \bibnamefont {Ceperley}},\
  }\bibfield  {title} {\bibinfo {title} {Liquid\&\#x2013;liquid phase
  transition in hydrogen by coupled electron\&\#x2013;ion monte carlo
  simulations},\ }\href {https://doi.org/10.1073/pnas.1603853113} {\bibfield
  {journal} {\bibinfo  {journal} {Proc. Natl. Acad. Sci.}\ }\textbf {\bibinfo
  {volume} {113}},\ \bibinfo {pages} {4953} (\bibinfo {year}
  {2016})}\BibitemShut {NoStop}%
\bibitem [{\citenamefont {Geng}\ \emph {et~al.}(2019)\citenamefont {Geng},
  \citenamefont {Wu}, \citenamefont {Marqu\'es},\ and\ \citenamefont
  {Ackland}}]{Geng_H_2019}%
  \BibitemOpen
  \bibfield  {author} {\bibinfo {author} {\bibfnamefont {H.~Y.}\ \bibnamefont
  {Geng}}, \bibinfo {author} {\bibfnamefont {Q.}~\bibnamefont {Wu}}, \bibinfo
  {author} {\bibfnamefont {M.}~\bibnamefont {Marqu\'es}},\ and\ \bibinfo
  {author} {\bibfnamefont {G.~J.}\ \bibnamefont {Ackland}},\ }\bibfield
  {title} {\bibinfo {title} {Thermodynamic anomalies and three distinct
  liquid-liquid transitions in warm dense liquid hydrogen},\ }\href
  {https://doi.org/10.1103/PhysRevB.100.134109} {\bibfield  {journal} {\bibinfo
   {journal} {Phys. Rev. B}\ }\textbf {\bibinfo {volume} {100}},\ \bibinfo
  {pages} {134109} (\bibinfo {year} {2019})}\BibitemShut {NoStop}%
\bibitem [{\citenamefont {Hinz}\ \emph {et~al.}(2020)\citenamefont {Hinz},
  \citenamefont {Karasiev}, \citenamefont {Hu}, \citenamefont {Zaghoo},
  \citenamefont {Mej\'{\i}a-Rodr\'{\i}guez}, \citenamefont {Trickey},\ and\
  \citenamefont {Calder\'{\i}n}}]{Heinz_H_2020}%
  \BibitemOpen
  \bibfield  {author} {\bibinfo {author} {\bibfnamefont {J.}~\bibnamefont
  {Hinz}}, \bibinfo {author} {\bibfnamefont {V.~V.}\ \bibnamefont {Karasiev}},
  \bibinfo {author} {\bibfnamefont {S.~X.}\ \bibnamefont {Hu}}, \bibinfo
  {author} {\bibfnamefont {M.}~\bibnamefont {Zaghoo}}, \bibinfo {author}
  {\bibfnamefont {D.}~\bibnamefont {Mej\'{\i}a-Rodr\'{\i}guez}}, \bibinfo
  {author} {\bibfnamefont {S.~B.}\ \bibnamefont {Trickey}},\ and\ \bibinfo
  {author} {\bibfnamefont {L.}~\bibnamefont {Calder\'{\i}n}},\ }\bibfield
  {title} {\bibinfo {title} {Fully consistent density functional theory
  determination of the insulator-metal transition boundary in warm dense
  hydrogen},\ }\href {https://doi.org/10.1103/PhysRevResearch.2.032065}
  {\bibfield  {journal} {\bibinfo  {journal} {Phys. Rev. Research}\ }\textbf
  {\bibinfo {volume} {2}},\ \bibinfo {pages} {032065} (\bibinfo {year}
  {2020})}\BibitemShut {NoStop}%
\bibitem [{\citenamefont {Cheng}\ \emph {et~al.}(2020)\citenamefont {Cheng},
  \citenamefont {Mazzola}, \citenamefont {Pickard},\ and\ \citenamefont
  {Ceriotti}}]{Cheng_H_2020}%
  \BibitemOpen
  \bibfield  {author} {\bibinfo {author} {\bibfnamefont {B.}~\bibnamefont
  {Cheng}}, \bibinfo {author} {\bibfnamefont {G.}~\bibnamefont {Mazzola}},
  \bibinfo {author} {\bibfnamefont {C.~J.}\ \bibnamefont {Pickard}},\ and\
  \bibinfo {author} {\bibfnamefont {M.}~\bibnamefont {Ceriotti}},\ }\bibfield
  {title} {\bibinfo {title} {Evidence for supercritical behaviour of
  high-pressure liquid hydrogen},\ }\href
  {https://doi.org/https://doi.org/10.1038/s41586-020-2677-y} {\bibfield
  {journal} {\bibinfo  {journal} {Nature}\ }\textbf {\bibinfo {volume} {585}},\
  \bibinfo {pages} {217} (\bibinfo {year} {2020})}\BibitemShut {NoStop}%
\bibitem [{\citenamefont {Goncharov}\ and\ \citenamefont
  {Geballe}(2017)}]{Comment_Zaghoo_2016_1}%
  \BibitemOpen
  \bibfield  {author} {\bibinfo {author} {\bibfnamefont {A.~F.}\ \bibnamefont
  {Goncharov}}\ and\ \bibinfo {author} {\bibfnamefont {Z.~M.}\ \bibnamefont
  {Geballe}},\ }\bibfield  {title} {\bibinfo {title} {Comment on ``evidence of
  a first-order phase transition to metallic hydrogen''},\ }\href
  {https://doi.org/10.1103/PhysRevB.96.157101} {\bibfield  {journal} {\bibinfo
  {journal} {Phys. Rev. B}\ }\textbf {\bibinfo {volume} {96}},\ \bibinfo
  {pages} {157101} (\bibinfo {year} {2017})}\BibitemShut {NoStop}%
\bibitem [{\citenamefont {Howie}\ \emph {et~al.}(2017)\citenamefont {Howie},
  \citenamefont {Dalladay-Simpson},\ and\ \citenamefont
  {Gregoryanz}}]{Comment_Zaghoo_2016_2}%
  \BibitemOpen
  \bibfield  {author} {\bibinfo {author} {\bibfnamefont {R.~T.}\ \bibnamefont
  {Howie}}, \bibinfo {author} {\bibfnamefont {P.}~\bibnamefont
  {Dalladay-Simpson}},\ and\ \bibinfo {author} {\bibfnamefont {E.}~\bibnamefont
  {Gregoryanz}},\ }\bibfield  {title} {\bibinfo {title} {Comment on ``evidence
  of a first-order phase transition to metallic hydrogen''},\ }\href
  {https://doi.org/10.1103/PhysRevB.96.157102} {\bibfield  {journal} {\bibinfo
  {journal} {Phys. Rev. B}\ }\textbf {\bibinfo {volume} {96}},\ \bibinfo
  {pages} {157102} (\bibinfo {year} {2017})}\BibitemShut {NoStop}%
\bibitem [{\citenamefont {Silvera}\ \emph {et~al.}(2017)\citenamefont
  {Silvera}, \citenamefont {Zaghoo},\ and\ \citenamefont
  {Salamat}}]{Reply_Comment_Zaghoo_2016}%
  \BibitemOpen
  \bibfield  {author} {\bibinfo {author} {\bibfnamefont {I.~F.}\ \bibnamefont
  {Silvera}}, \bibinfo {author} {\bibfnamefont {M.}~\bibnamefont {Zaghoo}},\
  and\ \bibinfo {author} {\bibfnamefont {A.}~\bibnamefont {Salamat}},\
  }\bibfield  {title} {\bibinfo {title} {Reply to ``comment on `evidence of a
  first-order phase transition to metallic hydrogen' ''},\ }\href
  {https://doi.org/10.1103/PhysRevB.96.237102} {\bibfield  {journal} {\bibinfo
  {journal} {Phys. Rev. B}\ }\textbf {\bibinfo {volume} {96}},\ \bibinfo
  {pages} {237102} (\bibinfo {year} {2017})}\BibitemShut {NoStop}%
\bibitem [{\citenamefont {Zaccarelli}\ \emph {et~al.}(2005)\citenamefont
  {Zaccarelli}, \citenamefont {Buldyrev}, \citenamefont {Nave}, \citenamefont
  {Moreno}, \citenamefont {Saika-Voivod}, \citenamefont {Sciortino},\ and\
  \citenamefont {Tartagliae}}]{Zaccarelli2005}%
  \BibitemOpen
  \bibfield  {author} {\bibinfo {author} {\bibfnamefont {E.}~\bibnamefont
  {Zaccarelli}}, \bibinfo {author} {\bibfnamefont {S.~V.}\ \bibnamefont
  {Buldyrev}}, \bibinfo {author} {\bibfnamefont {E.~L.}\ \bibnamefont {Nave}},
  \bibinfo {author} {\bibfnamefont {A.~J.}\ \bibnamefont {Moreno}}, \bibinfo
  {author} {\bibfnamefont {I.}~\bibnamefont {Saika-Voivod}}, \bibinfo {author}
  {\bibfnamefont {F.}~\bibnamefont {Sciortino}},\ and\ \bibinfo {author}
  {\bibfnamefont {P.}~\bibnamefont {Tartagliae}},\ }\bibfield  {title}
  {\bibinfo {title} {Model for reversible colloidal gelation},\ }\href
  {https://doi.org/10.1103/PhysRevLett.94.218301} {\bibfield  {journal}
  {\bibinfo  {journal} {Phys. Rev. Lett.}\ }\textbf {\bibinfo {volume} {94}},\
  \bibinfo {pages} {218301} (\bibinfo {year} {2005})}\BibitemShut {NoStop}%
\bibitem [{\citenamefont {Speedy}\ and\ \citenamefont
  {Debenedetti}(1994)}]{Debenedetti}%
  \BibitemOpen
  \bibfield  {author} {\bibinfo {author} {\bibfnamefont {R.~J.}\ \bibnamefont
  {Speedy}}\ and\ \bibinfo {author} {\bibfnamefont {P.~G.}\ \bibnamefont
  {Debenedetti}},\ }\bibfield  {title} {\bibinfo {title} {The entropy of a
  network crystal, fluid and glass},\ }\href
  {https://doi.org/10.1080/00268979400100161} {\bibfield  {journal} {\bibinfo
  {journal} {Mol. Phys.}\ }\textbf {\bibinfo {volume} {81}},\ \bibinfo {pages}
  {237} (\bibinfo {year} {1994})}\BibitemShut {NoStop}%
\bibitem [{\citenamefont {Speedy}\ and\ \citenamefont
  {Debenedetti}(1996)}]{Debenedetti2}%
  \BibitemOpen
  \bibfield  {author} {\bibinfo {author} {\bibfnamefont {R.~J.}\ \bibnamefont
  {Speedy}}\ and\ \bibinfo {author} {\bibfnamefont {P.~G.}\ \bibnamefont
  {Debenedetti}},\ }\bibfield  {title} {\bibinfo {title} {The distribution of
  tetravalent network glasses},\ }\href
  {https://doi.org/10.1080/00268979609484512} {\bibfield  {journal} {\bibinfo
  {journal} {Mol. Phys.}\ }\textbf {\bibinfo {volume} {88}},\ \bibinfo {pages}
  {1293} (\bibinfo {year} {1996})}\BibitemShut {NoStop}%
\bibitem [{\citenamefont {Stell}\ and\ \citenamefont
  {Hemmer}(1972)}]{Stell1972}%
  \BibitemOpen
  \bibfield  {author} {\bibinfo {author} {\bibfnamefont {G.}~\bibnamefont
  {Stell}}\ and\ \bibinfo {author} {\bibfnamefont {P.~C.}\ \bibnamefont
  {Hemmer}},\ }\bibfield  {title} {\bibinfo {title} {Phase transitions due to
  softness of the potential core},\ }\href
  {https://doi.org/https://doi.org/10.1063/1.1677857} {\bibfield  {journal}
  {\bibinfo  {journal} {J. Chem. Phys.}\ }\textbf {\bibinfo {volume} {56}},\
  \bibinfo {pages} {4274} (\bibinfo {year} {1972})}\BibitemShut {NoStop}%
\bibitem [{\citenamefont {Stillinger}\ and\ \citenamefont
  {Head-Gordon}(1993)}]{Stillinger1993}%
  \BibitemOpen
  \bibfield  {author} {\bibinfo {author} {\bibfnamefont {F.~H.}\ \bibnamefont
  {Stillinger}}\ and\ \bibinfo {author} {\bibfnamefont {T.}~\bibnamefont
  {Head-Gordon}},\ }\bibfield  {title} {\bibinfo {title} {Perturbational view
  of inherent structures in water},\ }\href
  {https://doi.org/10.1103/PhysRevE.47.2484} {\bibfield  {journal} {\bibinfo
  {journal} {Phys. Rev. E}\ }\textbf {\bibinfo {volume} {47}},\ \bibinfo
  {pages} {2484} (\bibinfo {year} {1993})}\BibitemShut {NoStop}%
\bibitem [{\citenamefont {Jagla}(2001)}]{Jagla2001}%
  \BibitemOpen
  \bibfield  {author} {\bibinfo {author} {\bibfnamefont {E.~A.}\ \bibnamefont
  {Jagla}},\ }\bibfield  {title} {\bibinfo {title} {Liquid-liquid equilibrium
  for monodisperse spherical particles},\ }\href
  {https://doi.org/10.1103/PhysRevE.63.061501} {\bibfield  {journal} {\bibinfo
  {journal} {Phys. Rev. E}\ }\textbf {\bibinfo {volume} {63}},\ \bibinfo
  {pages} {061501} (\bibinfo {year} {2001})}\BibitemShut {NoStop}%
\bibitem [{\citenamefont {Gibson}\ and\ \citenamefont
  {Wilding}(2006)}]{Gibson2006}%
  \BibitemOpen
  \bibfield  {author} {\bibinfo {author} {\bibfnamefont {H.~M.}\ \bibnamefont
  {Gibson}}\ and\ \bibinfo {author} {\bibfnamefont {N.~B.}\ \bibnamefont
  {Wilding}},\ }\bibfield  {title} {\bibinfo {title} {Metastable liquid-liquid
  coexistence and density anomalies in a core-softened fluid},\ }\href
  {https://doi.org/10.1103/PhysRevE.73.061507} {\bibfield  {journal} {\bibinfo
  {journal} {Phys. Rev. E}\ }\textbf {\bibinfo {volume} {73}},\ \bibinfo
  {pages} {061507} (\bibinfo {year} {2006})}\BibitemShut {NoStop}%
\bibitem [{\citenamefont {Skibinsky}\ \emph {et~al.}(2004)\citenamefont
  {Skibinsky}, \citenamefont {Buldyrev}, \citenamefont {Franzese},
  \citenamefont {Malescio},\ and\ \citenamefont {Stanley}}]{Skibinsky2004}%
  \BibitemOpen
  \bibfield  {author} {\bibinfo {author} {\bibfnamefont {A.}~\bibnamefont
  {Skibinsky}}, \bibinfo {author} {\bibfnamefont {S.~V.}\ \bibnamefont
  {Buldyrev}}, \bibinfo {author} {\bibfnamefont {G.}~\bibnamefont {Franzese}},
  \bibinfo {author} {\bibfnamefont {G.}~\bibnamefont {Malescio}},\ and\
  \bibinfo {author} {\bibfnamefont {H.~E.}\ \bibnamefont {Stanley}},\
  }\bibfield  {title} {\bibinfo {title} {Liquid-liquid phase transitions for
  soft-core attractive potentials},\ }\href
  {https://doi.org/10.1103/PhysRevE.69.061206} {\bibfield  {journal} {\bibinfo
  {journal} {Phys. Rev. E}\ }\textbf {\bibinfo {volume} {69}},\ \bibinfo
  {pages} {61206} (\bibinfo {year} {2004})}\BibitemShut {NoStop}%
\bibitem [{\citenamefont {Buldyrev}\ \emph {et~al.}(2010)\citenamefont
  {Buldyrev}, \citenamefont {Kumar}, \citenamefont {Sastry}, \citenamefont
  {Stanley},\ and\ \citenamefont {Weiner}}]{Buldyrev_Hydrophobic_2010}%
  \BibitemOpen
  \bibfield  {author} {\bibinfo {author} {\bibfnamefont {S.~V.}\ \bibnamefont
  {Buldyrev}}, \bibinfo {author} {\bibfnamefont {P.}~\bibnamefont {Kumar}},
  \bibinfo {author} {\bibfnamefont {S.}~\bibnamefont {Sastry}}, \bibinfo
  {author} {\bibfnamefont {H.~E.}\ \bibnamefont {Stanley}},\ and\ \bibinfo
  {author} {\bibfnamefont {S.}~\bibnamefont {Weiner}},\ }\bibfield  {title}
  {\bibinfo {title} {Hydrophobic collapse and cold denaturation in the jagla
  model of water},\ }\href@noop {} {\bibfield  {journal} {\bibinfo  {journal}
  {J. Phys.: Condens. Matter}\ }\textbf {\bibinfo {volume} {22}},\ \bibinfo
  {pages} {284109} (\bibinfo {year} {2010})}\BibitemShut {NoStop}%
\bibitem [{\citenamefont {Flory}(1941)}]{Flory_Polymer_1941}%
  \BibitemOpen
  \bibfield  {author} {\bibinfo {author} {\bibfnamefont {P.~J.}\ \bibnamefont
  {Flory}},\ }\bibfield  {title} {\bibinfo {title} {Thermodynamics of high
  polymer solutions},\ }\href
  {https://doi.org/https://doi.org/10.1063/1.1750971} {\bibfield  {journal}
  {\bibinfo  {journal} {J. Chem. Phys.}\ }\textbf {\bibinfo {volume} {9}},\
  \bibinfo {pages} {660} (\bibinfo {year} {1941})}\BibitemShut {NoStop}%
\bibitem [{\citenamefont {Huggins}(1941)}]{Huggins_Solutions_1941}%
  \BibitemOpen
  \bibfield  {author} {\bibinfo {author} {\bibfnamefont {M.~L.}\ \bibnamefont
  {Huggins}},\ }\bibfield  {title} {\bibinfo {title} {Solutions of long chain
  compounds},\ }\href {https://doi.org/https://doi.org/10.1063/1.1750930}
  {\bibfield  {journal} {\bibinfo  {journal} {J. Chem. Phys.}\ }\textbf
  {\bibinfo {volume} {9}},\ \bibinfo {pages} {440} (\bibinfo {year}
  {1941})}\BibitemShut {NoStop}%
\bibitem [{\citenamefont {Chan}\ and\ \citenamefont
  {Rey}(1996)}]{Rey_PIPS_1996}%
  \BibitemOpen
  \bibfield  {author} {\bibinfo {author} {\bibfnamefont {P.~K.}\ \bibnamefont
  {Chan}}\ and\ \bibinfo {author} {\bibfnamefont {A.~D.}\ \bibnamefont {Rey}},\
  }\bibfield  {title} {\bibinfo {title} {Polymerization-induced phase
  separation. 1. droplet size selection mechanism},\ }\href
  {https://doi.org/https://doi.org/10.1021/ma960690k} {\bibfield  {journal}
  {\bibinfo  {journal} {Macromolecules}\ }\textbf {\bibinfo {volume} {29}},\
  \bibinfo {pages} {8934} (\bibinfo {year} {1996})}\BibitemShut {NoStop}%
\bibitem [{\citenamefont {Luo}(2006)}]{Luo_PIPS_2006}%
  \BibitemOpen
  \bibfield  {author} {\bibinfo {author} {\bibfnamefont {K.}~\bibnamefont
  {Luo}},\ }\bibfield  {title} {\bibinfo {title} {The morphology and dynamics
  of polymerization-induced phase separation},\ }\href
  {https://doi.org/https://doi.org/10.1016/j.eurpolymj.2006.01.019} {\bibfield
  {journal} {\bibinfo  {journal} {European Polymer Journal}\ }\textbf {\bibinfo
  {volume} {42}},\ \bibinfo {pages} {1499} (\bibinfo {year}
  {2006})}\BibitemShut {NoStop}%
\bibitem [{\citenamefont {Hildebrand}\ and\ \citenamefont
  {Scott}(1962)}]{Hildebrand_Regular_1962}%
  \BibitemOpen
  \bibfield  {author} {\bibinfo {author} {\bibfnamefont {J.}~\bibnamefont
  {Hildebrand}}\ and\ \bibinfo {author} {\bibfnamefont {R.}~\bibnamefont
  {Scott}},\ }\href {https://books.google.com/books?id=fQUpAQAAMAAJ} {\emph
  {\bibinfo {title} {Regular Solutions}}},\ Prentice-Hall international series
  in chemistry\ (\bibinfo  {publisher} {Prentice-Hall},\ \bibinfo {year}
  {1962})\BibitemShut {NoStop}%
\bibitem [{\citenamefont {Alder}\ and\ \citenamefont
  {Wainwright}(1959)}]{Alder1959}%
  \BibitemOpen
  \bibfield  {author} {\bibinfo {author} {\bibfnamefont {B.~J.}\ \bibnamefont
  {Alder}}\ and\ \bibinfo {author} {\bibfnamefont {T.~E.}\ \bibnamefont
  {Wainwright}},\ }\bibfield  {title} {\bibinfo {title} {Studies in molecular
  dynamics. i. general method},\ }\href
  {https://doi.org/https://doi.org/10.1063/1.1730376} {\bibfield  {journal}
  {\bibinfo  {journal} {J. Chem. Phys.}\ }\textbf {\bibinfo {volume} {31}},\
  \bibinfo {pages} {459} (\bibinfo {year} {1959})}\BibitemShut {NoStop}%
\bibitem [{\citenamefont {Rapaport}(2004)}]{Rapaport2004}%
  \BibitemOpen
  \bibfield  {author} {\bibinfo {author} {\bibfnamefont {D.~C.}\ \bibnamefont
  {Rapaport}},\ }\href@noop {} {\emph {\bibinfo {title} {The Art of Molecular
  Dynamics Simulation}}},\ \bibinfo {edition} {2nd}\ ed.\ (\bibinfo
  {publisher} {Cambridge University Press},\ \bibinfo {address} {Cambridge,
  UK},\ \bibinfo {year} {2004})\BibitemShut {NoStop}%
\bibitem [{\citenamefont {Buldyrev}(2009)}]{Buldyrev_Application_2008}%
  \BibitemOpen
  \bibfield  {author} {\bibinfo {author} {\bibfnamefont {S.}~\bibnamefont
  {Buldyrev}},\ }\bibfield  {title} {\bibinfo {title} {Application of discrete
  molecular dynamics to protein folding and aggregation},\ }in\ \href
  {https://doi.org/https://doi.org/10.1007/978-3-540-78765-5_5} {\emph
  {\bibinfo {booktitle} {Aspects of Physical Biology}}},\ \bibinfo {series}
  {Lecture Notes in Physics}, Vol.\ \bibinfo {volume} {752},\ \bibinfo {editor}
  {edited by\ \bibinfo {editor} {\bibfnamefont {G.}~\bibnamefont {Franzese}}\
  and\ \bibinfo {editor} {\bibfnamefont {M.}~\bibnamefont {Rubi}}}\ (\bibinfo
  {publisher} {Springer-Verlag},\ \bibinfo {address} {Berlin, Heidelberg},\
  \bibinfo {year} {2009})\ pp.\ \bibinfo {pages} {97--132}\BibitemShut
  {NoStop}%
\bibitem [{\citenamefont {Berendsen}\ \emph {et~al.}(1984)\citenamefont
  {Berendsen}, \citenamefont {Postma}, \citenamefont {van Gunsteren},
  \citenamefont {DiNola},\ and\ \citenamefont {Haak}}]{Berendsen1984}%
  \BibitemOpen
  \bibfield  {author} {\bibinfo {author} {\bibfnamefont {H.~J.~C.}\
  \bibnamefont {Berendsen}}, \bibinfo {author} {\bibfnamefont {J.~P.~M.}\
  \bibnamefont {Postma}}, \bibinfo {author} {\bibfnamefont {W.~F.}\
  \bibnamefont {van Gunsteren}}, \bibinfo {author} {\bibfnamefont
  {A.}~\bibnamefont {DiNola}},\ and\ \bibinfo {author} {\bibfnamefont {J.~R.}\
  \bibnamefont {Haak}},\ }\bibfield  {title} {\bibinfo {title}
  {Molecular-dynamics with coupling to an external bath},\ }\href@noop {}
  {\bibfield  {journal} {\bibinfo  {journal} {J. Chem. Phys.}\ }\textbf
  {\bibinfo {volume} {81}},\ \bibinfo {pages} {3684} (\bibinfo {year}
  {1984})}\BibitemShut {NoStop}%
\bibitem [{\citenamefont {Luo}\ \emph {et~al.}(2015)\citenamefont {Luo},
  \citenamefont {Xu}, \citenamefont {Angell}, \citenamefont {Stanley},\ and\
  \citenamefont {Buldyrev}}]{Luo2015}%
  \BibitemOpen
  \bibfield  {author} {\bibinfo {author} {\bibfnamefont {J.}~\bibnamefont
  {Luo}}, \bibinfo {author} {\bibfnamefont {L.}~\bibnamefont {Xu}}, \bibinfo
  {author} {\bibfnamefont {C.~A.}\ \bibnamefont {Angell}}, \bibinfo {author}
  {\bibfnamefont {H.~E.}\ \bibnamefont {Stanley}},\ and\ \bibinfo {author}
  {\bibfnamefont {S.~V.}\ \bibnamefont {Buldyrev}},\ }\bibfield  {title}
  {\bibinfo {title} {Physics of the jagla model as the liquid-liquid
  coexistence line slope varies},\ }\href
  {https://doi.org/https://doi.org/10.1063/1.4921559} {\bibfield  {journal}
  {\bibinfo  {journal} {J. Chem. Phys.}\ }\textbf {\bibinfo {volume} {142}},\
  \bibinfo {pages} {224501} (\bibinfo {year} {2015})}\BibitemShut {NoStop}%
\bibitem [{\citenamefont {Xu}\ \emph {et~al.}(2005)\citenamefont {Xu},
  \citenamefont {Kumar}, \citenamefont {Buldyrev}, \citenamefont {Chen},
  \citenamefont {Poole}, \citenamefont {Sciortino},\ and\ \citenamefont
  {Stanley}}]{Xu_LLPT_2005}%
  \BibitemOpen
  \bibfield  {author} {\bibinfo {author} {\bibfnamefont {L.}~\bibnamefont
  {Xu}}, \bibinfo {author} {\bibfnamefont {P.}~\bibnamefont {Kumar}}, \bibinfo
  {author} {\bibfnamefont {S.~V.}\ \bibnamefont {Buldyrev}}, \bibinfo {author}
  {\bibfnamefont {S.-H.}\ \bibnamefont {Chen}}, \bibinfo {author}
  {\bibfnamefont {P.~H.}\ \bibnamefont {Poole}}, \bibinfo {author}
  {\bibfnamefont {F.}~\bibnamefont {Sciortino}},\ and\ \bibinfo {author}
  {\bibfnamefont {H.~E.}\ \bibnamefont {Stanley}},\ }\bibfield  {title}
  {\bibinfo {title} {Relation between the widom line and the dynamic crossover
  in systems with a liquid–liquid phase transition},\ }\href
  {https://doi.org/10.1073/pnas.0507870102} {\bibfield  {journal} {\bibinfo
  {journal} {Proc. Natl. Acad. Sci.}\ }\textbf {\bibinfo {volume} {102}},\
  \bibinfo {pages} {16558} (\bibinfo {year} {2005})}\BibitemShut {NoStop}%
\bibitem [{\citenamefont {Lide}(2003)}]{Lide2003}%
  \BibitemOpen
  \bibfield  {author} {\bibinfo {author} {\bibfnamefont {D.~R.}\ \bibnamefont
  {Lide}},\ }\href@noop {} {\emph {\bibinfo {title} {CRC Handbook of Chemistry
  and Physics}}},\ \bibinfo {edition} {85th}\ ed.\ (\bibinfo  {publisher} {CRC
  Press},\ \bibinfo {address} {Boca Raton, FL},\ \bibinfo {year}
  {2003})\BibitemShut {NoStop}%
\end{thebibliography}%

\end{document}